\documentclass[11pt,a4paper,nofootinbib,showpacs,preprint,utf8]{article}
\usepackage{jheppub}
\pdfoutput=1
\usepackage{amsmath,amssymb}
\usepackage{amsfonts}
\usepackage{epsfig}
\usepackage{graphicx}
\usepackage{amssymb}
\usepackage{bbm}
\usepackage{pifont}
\usepackage[usenames,dvipsnames]{xcolor}
\definecolor{warmblack}{rgb}{0.0, 0.26, 0.26}
\definecolor{mediumtealblue}{rgb}{0.0, 0.33, 0.71}
\usepackage[compat=1.1.0]{tikz-feynman}
\usepackage{tikzsymbols}
\usepackage{comment}
\usepackage[normalem]{ulem}
\usepackage{orcidlink}

\usepackage{dcolumn}
\usepackage{bm}
\usepackage{graphicx}
\usepackage{amssymb,amsmath}
\usepackage{multirow}
\usepackage{afterpage}
\usepackage{bbm}
\usepackage{enumerate}   
\PassOptionsToPackage{colorlinks=true
,urlcolor=magenta
,anchorcolor=blue
,citecolor=blue
,filecolor=blue
,linkcolor=blue
,menucolor=blue
,linktocpage=true
,pdfproducer=medialab
,pdfa=true
}{hyperref}

\usepackage{tikz-feynman}
\tikzfeynmanset{compat=1.1.0}
\usetikzlibrary{trees}
\usetikzlibrary{decorations.pathmorphing}
\usetikzlibrary{decorations.markings}

\usepackage{array,booktabs}

\usepackage{slashed}
\usepackage{epsfig,psfrag,rotating,soul}
\usepackage{rotfloat}

\allowdisplaybreaks

\newcommand{\beq}{\begin{equation}}
\newcommand{\eeq}{\end{equation}}
\newcommand{\clphi}{c^{}_{\text{L}\Phi}}
\newcommand{\crphi}{c^{}_{\text{R}\Phi}}
\newcommand{\clchi}{c^{}_{\text{L}\chi}}
\newcommand{\crchi}{c^{}_{\text{R}\chi}}
\newcommand{\gev}{{\text{GeV}}}
\newcommand{\trh}{T_{\text{rh}}}
\newcommand{\tev}{{\text{TeV}}}

\usepackage{xspace}

\newcommand{\Eq}{Eq\,.~}
\newcommand{\Eqs}{Eqs\,.~}
\newcommand{\Fig}{Fig\,.~}
\newcommand{\gss}{g_{\star s}}
\newcommand{\Trh}{T_\text{rh}}
\newcommand{\arh}{a_\text{rh}}
\newcommand{\Tmax}{T_\text{max}}

\newcommand{\rR}{\rho_R}

\newcommand{\Gp}{\Gamma_\phi}

\newcommand{\ndm}{n_{\rm DM}}
\newcommand{\Ndm}{N_{\rm DM}}
\newcommand{\mdm}{m_{\rm DM}}
\newcommand{\mphi}{m_{\rm \Phi}}
\newcommand{\mchi}{m_{\chi}}
\newcommand{\adm}{a_{\rm DM}}
\newcommand{\lI}{\Lambda_{\rm I}}
\newcommand{\lNP}{\Lambda_{\rm NP}}
\newcommand{\hub}{\mathcal{H}}
\newcommand{\rhophi}{\rho_{\phi}}
\newcommand{\rhoR}{\rho_{R}}

\newcommand{\ba}{\begin{array}}
\newcommand{\ea}{\end{array}}
\newcommand{\bd}{\begin{displaymath}}
\newcommand{\ed}{\end{displaymath}}
\newcommand{\besub}{\begin{subequations}}
\newcommand{\eesub}{\end{subequations}}
\newcommand{\bea}{\begin{eqnarray}}
\newcommand{\eea}{\end{eqnarray}}

\def\q2 {q^2}

\def\bt{\begin{table}}
\def\et{\end{table}}

\tikzfeynmanset{every dot/.style={minimum size=5pt}}

\definecolor{mygray}{gray}{0.85} 
\definecolor{myblue}{cmyk}{0.65, 0.37, 0.0, 0.19}

\title{Multi-messenger FIMP}

\author[a,b]{{Debasish Borah,}}
\author[a]{{Nayan Das,}}
\author[c,d]{{Sahabub Jahedi,}}
\author[a]{{and Dipankar Pradhan}}

\affiliation[a]{Department of Physics, Indian Institute of Technology Guwahati, Assam 781039, India}
\affiliation[b]{Pittsburgh Particle Physics, Astrophysics, and Cosmology Center, Department of Physics and Astronomy, University of Pittsburgh, Pittsburgh, PA 15260, USA}
\affiliation[c]{State Key Laboratory of Nuclear Physics and Technology, Institute of Quantum Matter, South China Normal University, Guangzhou 510006, China}
\affiliation[d]{Guangdong Basic Research Center of Excellence for Structure and Fundamental Interactions of Matter, Guangdong Provincial Key Laboratory of Nuclear Science, Guangzhou 510006, China}

\emailAdd{dborah@iitg.ac.in}
\emailAdd{nayan.das@iitg.ac.in}
\emailAdd{sahabub@m.scnu.edu.cn}
\emailAdd{d.pradhan@iitg.ac.in}

\abstract{We propose a multi-messenger frontier probe of non-thermal or freeze-in massive particle (FIMP) dark matter (DM) by considering an effective field theory (EFT) setup. Assuming leptophilic operators connecting DM with the standard model (SM) bath, we consider DM mass ($m_{\rm DM}$) and the reheat temperature of the Universe ($T_{\rm rh}$) in a regime which prevents DM-SM thermalisation. Low $T_{\rm rh}$ allows sizeable DM-SM interactions even for non-thermal DM allowing the latter to be probed at direct, indirect detection frontiers as well as future electron-positron and muon colliders. An extended reheating period governed by monomial inflaton potential after its slow-roll phase not only generates the required abundance of non-thermal DM via ultraviolet (UV) freeze-in but also brings the scale-invariant primordial gravitational waves (GW) within reach of near future experiments across a wide range of frequencies. While particle physics experiments can probe $T_{\rm rh} \sim \mathcal{O}(10)$ GeV and FIMP DM with mass $m_{\rm DM} \sim \mathcal{O}(1)$ TeV, future GW detectors are sensitive to a much wider parameter space.}

\makeatletter
\gdef\@fpheader{}
\makeatother

\begin{document}
\maketitle

\section{Introduction}
\label{sec:intro}
We live in a Universe whose matter content at present is dominated by dark matter (DM) leaving only $\sim 20\%$ of the matter density to be composed of ordinary or visible matter \cite{Planck:2018vyg}. Evidences from cosmology and astrophysics based observations suggest DM to behave like pressureless, non-luminous and non-baryonic form of matter \cite{Planck:2018vyg, ParticleDataGroup:2024cfk, Cirelli:2024ssz}. Quantitatively, the present abundance of DM is quoted in terms of the respective density parameter $\Omega_{\rm DM}$ and reduced Hubble constant $h = \text{Hubble Parameter}/(100 \;\text{km} ~\text{s}^{-1} 
\text{Mpc}^{-1})$ as \cite{Planck:2018vyg}
\begin{equation}
\Omega_{\text{DM}} h^2 = 0.120\pm 0.001
\label{dm_relic}
\end{equation}
\noindent at 68\% CL. While none of the standard model (SM) particles can give rise to the observed DM abundance, weakly interacting massive particle (WIMP) \cite{Kolb:1990vq, Jungman:1995df, Bertone:2004pz} has emerged as the leading beyond standard model (BSM) paradigm for particle DM. The most appealing feature of WIMP is its mass and interactions around the electroweak scale ballpark which naturally leads to the observed DM relic after thermal freeze-out. Due to sizeable non-gravitational DM-SM interactions in WIMP scenario, it is also possible to detect it and terrestrial laboratories via DM-nucleon \cite{LUX-ZEPLIN:2022qhg,LZ:2024} or DM-electron scatterings \cite{XENON:2022ltv}. However, no such DM-SM interactions have been observed so far pushing the WIMP parameter space into a tight corner. While such null results at direct detection experiments do not rule out the WIMP hypothesis completely, it has motivated the particle physics community to look for alternative scenarios like freeze-in or feebly interacting massive particle (FIMP) DM
\cite{McDonald:2001vt, Hall:2009bx, Blennow:2013jba, Klasen:2013ypa, Elahi:2014fsa, Biswas:2016bfo, Biswas:2018aib, Barman:2020plp, Barman:2021tgt, Belanger:2020npe}
where DM, due to its feeble interactions with the SM bath, never enters equilibrium in the early universe. Instead of WIMP type freeze-out, such DM freezes in from the SM bath either via decay or scattering. Due to the non-thermal origin, such DM can carry signatures of early Universe history which typical thermal DM remains insensitive to thereby providing an interesting way to probe or constrain physics of the early Universe. On the other hand, such DM typically have feeble couplings with the SM, making direct detection rates negligible \footnote{See \cite{Hambye:2018dpi, Belanger:2018sti, Elor:2021swj, Bhattiprolu:2023akk} for some specific scenarios with detection possibilities.}. A recent review of such models can be found in \cite{Bernal:2017kxu}.

In this work, we consider a freeze-in scenario where thermal production of DM is prevented by a low reheat temperature $T_{\rm rh}$ after inflation. In these scenarios, DM production can occur via freeze-in even with sizeable DM-SM interactions, leading to interesting detection prospects. This possibility has been explored recently in several works \cite{Bhattiprolu:2022sdd, Boddy:2024vgt, Bringmann:2021sth, Cosme:2023xpa, Cosme:2024ndc, Arcadi:2024wwg, Barman:2024nhr,  Arcadi:2024obp, Barman:2024tjt, Bernal:2024ndy, Lee:2024wes, Belanger:2024yoj, Khan:2025keb} with the details of production mechanisms or detection prospects at direct search, collider experiments within specific model frameworks. Here we provide a complete picture of such detection aspects of freeze-in DM at multi-messenger avenues by adopting a model-independent effective field theory (EFT) setup. We consider both scalar and fermionic DM coupling to the SM bath via dimension-6 operators. For simplicity, we consider DM to be leptophilic such that the number of operators remain limited. EFT of DM has been studied in several works \cite{Beltran:2008xg, Fan:2010gt, Goodman:2010ku, Beltran:2010ww, Fitzpatrick:2012ix, Bertuzzo:2017lwt} in the context of direct detection, indirect detection as well as collider searches, also summarised in a recent review \cite{Bhattacharya:2021edh}. While leptophilic nature of DM helps in evading strong direct-detection bounds \cite{LZ:2022lsv,LZ:2024} to some extent, it also opens up interesting discovery prospects of DM at lepton colliders like the $e^+ e^-$ colliders \cite{Fox:2011fx, Essig:2013vha, Yu:2013aca, Kadota:2014mea, Yu:2014ula, Freitas:2014jla, Dutta:2017ljq, Liu:2019ogn, Choudhury:2019sxt, Kundu:2021cmo, Barman:2021hhg, Bhattacharya:2022wtr, Bhattacharya:2022qck, Ge:2023wye, Roy:2024ear, Borah:2024twm} as well as muon colliders \cite{Han:2020uak}.

In our setup DM relic is generated via ultraviolet (UV) freeze-in \cite{Hall:2009bx, Elahi:2014fsa,Barman:2020plp} occurring after the end of slow-roll inflation. Within the regime of perturbative reheating, we consider different monomial potential of the inflaton field during reheating era followed by the decay of the inflaton condensate either via bosonic or fermionic decay modes to reheat the Universe. Such non-trivial reheating era can lead to a bath temperature much more than the reheat temperature $\Tmax \gg \Trh$. DM yield via UV freeze-in is very sensitive to the maximum temperature, $\Tmax$, reached by the SM plasma~\cite{Giudice:2000ex, Garcia:2020eof} prior to the reheating. Due to sizeable DM-SM coupling, such DM can leave signatures at conventional direct and indirect search experiments with interesting detection aspects at future lepton colliders via mono-photon $+$ missing energy searches. On the other hand, the stiff equation of state during the reheating era can bring the primordial gravitational waves (GW) spectrum within reach of future detectors\footnote{Recently, the authors of \cite{Konar:2025iuk} studied high-frequency GW signatures of IR freeze-in scenario with graviton bremsstrahlung from one of the external legs.}. We find interesting complementarity among particle physics experiments and future GW detectors in probing the parameter space of the EFT cut-off scale $\lNP$ and DM mass $m_{\rm DM}$. While particle physics experiments like colliders can probe $\lNP \lesssim 10 \, {\rm TeV}, m_{\rm DM} \lesssim \mathcal{O}(1) \, {\rm TeV}$, future GW experiments can probe a much wider parameter space in $\lNP-m_{\rm DM}$ parameter space for such FIMP DM.


This paper is organised as follows. In section \ref{sec1}, we list the effective DM-SM operators considered in our analysis. In section \ref{sec2} we discuss the post-inflationary dynamics and production of DM via UV freeze-in. In section \ref{sec3} we discuss constraints and discovery prospects of DM. Finally, we summarise our results and conclude in section \ref{sec4}.

\section{Effective Operators}
\label{sec1}
As mentioned before, we consider both scalar ($\Phi$) and fermion ($\chi$) DM in our setup. While DM is a singlet under SM gauge symmetries, it is odd under an unbroken $\mathbbm{Z}_2$ symmetry, guaranteeing its stability. We further assume the DM to couple only with the SM leptons for simplicity. While this limits the detection prospects at typical direct search or hadron collider experiments, there exist plenty of opportunities via electron recoil, future lepton collider experiments, among others, as we discuss in the remainder of this work. With these assumptions, the dimension-6 operators invariant under the SM gauge and $\mathbbm{Z}_2$ symmetries connecting leptophilic DM with the SM are 
\begin{align}\label{eq:DM-model1}
\mathcal{O}^{V}_{L\Phi}=&\dfrac{c_{L \Phi}^{V}}{\lNP^2}(\bar{L}\gamma^{\mu} L)(\Phi^{\dagger}\overleftrightarrow{\partial_{\mu}} \Phi)\,,~(\times)\\
\mathcal{O}^{V}_{e\Phi}=&\dfrac{c_{e \Phi}^{V}}{\lNP^2}(\bar{e}\gamma^{\mu} e)(\Phi^{\dagger}\overleftrightarrow{\partial_{\mu}} \Phi)\,,~(\times)\\
\mathcal{O}^{V}_{L\chi}=&\dfrac{c_{L \chi}^{V}}{\lNP^2}(\bar{L}\gamma^{\mu} L) (\bar{\chi} \gamma_{\mu} \chi)\,,~(\times)\\
\mathcal{O}^{V}_{e\chi}=&\dfrac{c_{e \chi}^{V}}{\lNP^2}(\bar{e}\gamma^{\mu} e) (\bar{\chi} \gamma_{\mu} \chi)\,,~(\times)\\
 \label{eq:DM-model4}
 \mathcal{O}^{A}_{L\chi}=&\dfrac{c_{L \chi}^{A}}{\lNP^2}(\bar{L}\gamma^{\mu} L) (\bar{\chi} \gamma_{\mu} \gamma^5 \chi)\,,\\
 \mathcal{O}^{A}_{e\chi}=&\dfrac{c_{e \chi}^{A}}{\lNP^2}(\bar{e}\gamma^{\mu} e) (\bar{\chi} \gamma_{\mu} \gamma^5 \chi)\,,
 \label{eq:DM-model4}
\end{align}
where $L(e)$ is the $SU(2)_L$ doublet (singlet) leptons and $\lNP$ is the cut-off scale. For two fields $A$ and $B$, the bidirectional derivative is defined as $A\overleftrightarrow{\partial_{\mu}} B=A(\partial_{\mu} B)-(\partial_{\mu} A)B$. The symbol `$(\times)$' indicates that the relevant operator vanishes for real (Majorana) scalar (fermionic) DM. For the complex scalar DM scenario, we democratically set $c_{L \Phi}^{V} = c_{e \Phi}^{V} = 1$. In the case of fermionic DM, we focus exclusively on the vector DM current. Inclusion of the axial-vector DM current would approximately double the total cross-section, leading to a mild improvement in the phenomenological constraints on $\lNP$ by a factor of about 1.19. Therefore, to maintain a simplified framework, we consider Dirac fermionic DM having vector-like couplings only with $c_{L \chi}^{V} = c_{e \chi}^{V} = 1$ in our analysis. While we do not mention lepton flavor index explicitly in our operators, we consider flavor universal DM-SM couplings unless otherwise mentioned. For both scalar and fermionic DM, we will specify the values of the Wilson coefficients for vector-like couplings only without explicitly specifying them. It should be noted that, similar to the SM, we also have baryon and lepton number conservation in the DM-SM operators.

\section{Reheating Dynamics and UV Freeze-in of DM}
\label{sec2}
We elaborate on the freeze-in production of DM during reheating after inflation. We begin by examining the post-inflationary dynamics of the inflaton, which leads to the generation of a radiation bath that subsequently serves as the source for DM production. At the end of inflation, oscillation of the inflaton occurs at the bottom of a potential $V(\phi)$ parametrised by \cite{Garcia:2020wiy}
\begin{equation}\label{eq:inf-pot}
V(\phi) = \lambda\, \frac{\phi^n}{\Lambda_{\rm I}^{n - 4}},
\end{equation}
where $\lambda$ is a dimensionless coupling, and $\lI$ is the inflationary scale, constrained by CMB observations to be $\lI \lesssim 10^{16}$ GeV \cite{Planck:2018jri}. This potential naturally arises in several inflationary models, such as the $\alpha$-attractor T- and E-models \cite{Kallosh:2013hoa, Kallosh:2013maa,Kallosh:2013yoa} and the Starobinsky model \cite{Starobinsky:1980te,Starobinsky:1983zz, Kofman:1985aw}. Notably, the scale $\Lambda_{\rm I}$ governing inflation is distinct from $\lNP$, which characterizes DM-SM interactions, and the two can differ significantly.

The effective mass of the inflaton $m_\phi(a)$ is obtained from the second derivative of \Eq\eqref{eq:inf-pot}:
\begin{equation}\label{eq:inf-mass1}
m^2_\phi(a) = n(n - 1)\lambda \frac{\phi^{n - 2}}{\Lambda^{n - 4}_{\rm I}} \simeq n(n-1)\lambda^{2/n}\Lambda^{2(4 - n)/n}_{\rm I} \rho^{(n-2)/n}_\phi(a).
\end{equation}
For $n \neq 2$, $m_{\phi}$ depends on the field value, leading to a time-dependent inflaton decay rate. The equation of motion for the inflaton field is given by \cite{Turner:1983he}
\begin{equation}\label{eq:eom0}
\ddot\phi + (3 \hub + \Gamma_\phi) \dot\phi + V'(\phi) = 0,
\end{equation}
where $\hub$ is the Hubble expansion rate, $\Gamma_{\phi}$ is the inflaton decay rate, while the dots and primes denote derivatives with respect to time $t$ and field $\phi$, respectively. The evolution of the inflaton energy density, \(\rhophi = \frac{1}{2} \dot\phi^2 + V(\phi)\), is governed by the Boltzmann equation
\begin{equation}\label{eq:drhodt}
\frac{d\rhophi}{dt} + \frac{6n}{2 + n} \hub \rhophi = - \frac{2n}{2 + n} \Gamma_{\phi} \rhophi.
\end{equation}
where $\hub=\sqrt{(\rhophi+\rhoR)/(3M_P^2)}$ is the Hubble rate and $M_P=2.435\times 10^{18}~\gev$ is the reduced Planck mass. However, the inflaton oscillation behaves like a fluid with an effective equation of state $\omega\equiv p_\phi/\rhophi=(n-2)/(n+2)$ where $p_\phi\equiv \dfrac{1}{2}\dot{\phi}^2-V(\phi)$ is the pressure of the inflaton field. In Eq\,.~\eqref{eq:drhodt}, the term $\Gamma_{\phi}\rhophi$ accounts for the energy density transfer from inflaton to radiation via decays, and the term $\hub \rhophi$ accounts for the redshift dilution due to the expansion of the Universe. However, during reheating, when $a_I \ll a \ll a_{\rm rh}$, the expansion term $\hub \rho_\phi$ dominates over the decay term $\Gamma_\phi \rhophi$ and this allows an analytical solution:
\begin{equation}\label{eq:rpsol}
\rhophi(a) \simeq \rhophi (a_{\rm rh}) \left(\frac{a_{\rm rh}}{a}\right)^{6n/(2 + n)}.
\end{equation}
Since \(\hub\) is primarily determined by \(\rhophi\) during reheating, we find
\begin{equation}\label{eq:Hubble}
\hub(a) \simeq \hub(a_{\rm rh}) \times \begin{cases}
\left(\frac{a_{\rm rh}}{a}\right)^{3n/(n + 2)} &\text{for } a \leq a_{\rm rh},\\[10pt]
\left(\frac{a_{\rm rh}}{a}\right)^2 &\text{for } a_{\rm rh} \leq a.
\end{cases}
\end{equation}
Concurrently, the radiation energy density, $\rho_R$, evolves according to the continuity equation \cite{Garcia:2020wiy}:
\begin{equation}\label{eq:rR}
\frac{d\rhoR}{dt} + 4 \hub \rhoR = \frac{2n}{2 + n} \Gamma_\phi \rhophi.
\end{equation}
Reheating concludes when $\rho_R$ equals $\rho_\phi$:
\begin{equation}
\rho_R(a_{\rm rh}) = \rho_\phi(a_{\rm rh}) = 3 M_P^2 \hub(a_{\rm rh})^2.
\label{eq:rad}
\end{equation}
To maintain consistency with Big Bang Nucleosynthesis (BBN), the reheating temperature must satisfy $T_{\rm rh} > T_{\rm BBN} \simeq 4$ MeV \cite{Sarkar:1995dd, Kawasaki:2000en}.

\subsection{Reheating: fermionic and bosonic}
Following the end of inflation, the inflaton field typically undergoes coherent oscillations near the minimum of its potential. These oscillations result in the perturbative decay of the inflaton into SM particles, which subsequently interact and thermalize, giving rise to a hot, thermalized plasma. We consider two-body decay processes enabled by trilinear interactions between the inflaton $\phi$ and a pair of either complex scalar doublet $\varphi$ ($e.g.$, the Higgs doublet) or vector-like Dirac fermion $\psi$. The relevant interaction terms in the Lagrangian are given by
\begin{equation}
\mathcal{L}_{\text{int}} \supset -\mu_{\varphi}\, \phi\, |\varphi|^2 - y_\psi\, \bar{\psi} \psi\, \phi,
\label{eq:lag.infla}
\end{equation}
where $\mu_{\varphi}$ and $y_{\psi}$ are the trilinear scalar coupling of mass dimension 1 and Yukawa coupling of mass dimension 0, respectively. In this setup, we assume that the inflaton decays entirely into visible sector particles that form the radiation bath. We remain agnostic about the UV completion of the coupling between the inflaton and SM fields, maintaining a model-independent approach. Throughout our analysis, we consider each decay channel separately as an individual reheating scenario, with the assumption that multiple decay modes do not occur simultaneously.

If the inflaton decays exclusively into $\psi\bar{\psi}$ pair following the Yukawa interaction written in \Eq\eqref{eq:lag.infla}, the corresponding decay rate is given by
\begin{equation}
\Gamma_{\psi}(a)=\frac{y^2_{\text{eff}}}{8\pi}m_{\phi},~~~\text{with}~~~y_{\text{eff}}= \sqrt{\frac{8 \pi \hub(\arh)}{m_{\phi}(\arh)}};
\end{equation}
where the effective coupling $y_{\text{eff}} \neq y_{\psi}$ (for $n\neq2$) is determined after averaging over oscillations. Therefore, following \Eq\eqref{eq:rad}, the evolution of radiation energy density follows \cite{Bernal:2022wck}
\begin{equation} \label{eq:rR_fer}
\rhoR(a) \simeq \frac{3\, n}{7 - n}\, M_P^2\, \Gamma_{\psi}(\arh)\, \hub(\arh) \left(\frac{\arh}{a}\right)^\frac{6 (n - 1)}{2 + n} \left[1 - \left(\frac{a_I}{a}\right)^\frac{2 (7 - n)}{2 + n}\right]\,.
\end{equation}
The corresponding temperature of the thermal bath takes the form
\begin{equation}
T(a) \simeq \Trh \left(\frac{\arh}{a}\right)^\alpha,
\label{eq:Tevol}
\end{equation}
with
\begin{equation}
\alpha =
\begin{cases}
\frac32\, \frac{n - 1}{n + 2} & \text{ for } n < 7\,,\\
1 & \text{ for } n > 7\,.
\end{cases}
\label{eq:Tfer}
\end{equation}
By expressing the scale factor in terms of temperature, the Hubble parameter during the reheating phase (see \Eq\eqref{eq:Hubble}) can be recast as
\begin{equation}
\hub(T) \simeq \hub(T_{\text{rh}}) \left( \dfrac{T}{\Trh} \right)^{\frac{3}{\alpha}\frac{n}{n+2 } \,} \,,
\label{eq:Hevol}
\end{equation}
where $T_{\text{rh}}$ denotes the reheating temperature.

When the inflaton decays predominantly into a pair of bosons via the trilinear scalar interaction specified in \Eq\eqref{eq:lag.infla}, the decay rate is given by
\begin{equation}
\Gamma_\phi^{(a)} = \frac{\mu_{\text{eff}}^2}{8\pi m_{\phi}{(a)}},~~~\text{with}~~~\mu_{\text{eff}}=\sqrt{8 \pi \hub(\arh)m_{\phi}(\arh)}.
\end{equation}
Here, the effective coupling $\mu_{\text{eff}}\neq \mu$ (if $n \neq 2$) can be computed after averaging over oscillations. Similar to the fermionic reheating case, the radiation energy density scales as \cite{Bernal:2022wck}
\begin{equation}
\rR(a) \simeq \frac{3\, n}{1 + 2\, n}\, M_P^2\, \Gp(\arh)\, \hub(\arh) \left(\frac{\arh}{a}\right)^\frac{6}{2 + n} \left[1 - \left(\frac{a_I}{a}\right)^\frac{2\, (1 + 2 n)}{2 + n}\right],  
\label{eq:rR_bos}
\end{equation}
with which the SM temperature and the Hubble expansion rate evolve in accordance with Eqs.~\eqref{eq:Tevol} and~\eqref{eq:Hevol}, respectively, with
\begin{equation}
\alpha = \frac{3}{2(n + 2)}\,
\end{equation}
during reheating. For $n = 2$, the inflation equation of state behaves as non-relativistic matter ($w = 0$). Substituting this into \Eq\eqref{eq:Tevol}, we reproduce the standard dependence of temperature and the scale factor for an inflaton oscillating in a quadratic potential.
\begin{figure}[t]
\centering
\includegraphics[width=0.75\linewidth]{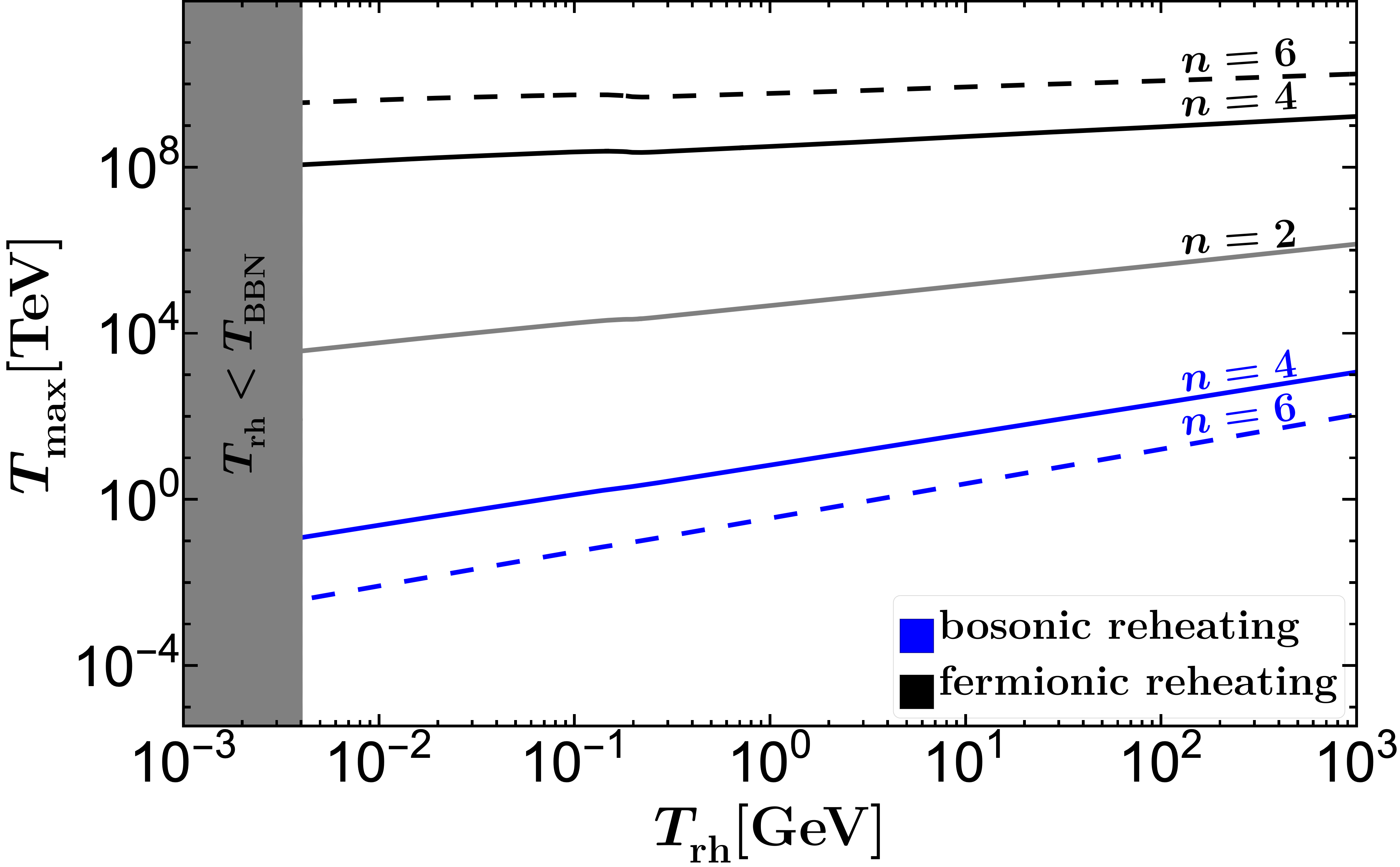}
\caption{The blue and black lines, shown as thick ($n=4$) and dashed ($n=6$), represent the bosonic and fermionic reheating scenarios, respectively. The thick gray line, common for both bosonic and fermionic reheating, corresponds to the $n=2$ scenario. The dark gray shaded region is excluded by BBN constraints. The parameters $\mu_{\rm eff}$, $y_{\rm eff}$ and $m_{\phi}(a_{\rm rh})$ are fixed by values of $n$ and $T_{\rm rh}$ for a particular number of e-fold during inflation. For example, for $T_{\rm rh}= 1$ GeV, $n=2$ gives $\mu_{\rm eff} \simeq 2.5\times10^{-2} $ GeV,  $y_{\rm eff} \simeq 1.6\times10^{-15} $ and $m_{\phi}(a_{\rm rh}) \simeq 1.5\times 10^{13}$ GeV. For the same reheating temperature, $n=6$ gives $\mu_{\rm eff} \simeq 2.1\times10^{-12} $ GeV, $y_{\rm eff} \simeq 2.0\times 10^{-5} $ and $m_{\phi}(a_{\rm rh}) \simeq 1.1\times 10^{-7}$ GeV. Here we fix the number of e-folds during inflation to $55$.  }
\label{fig:Trh-Tmax}
\end{figure}

Furthermore, the maximum temperature reached by the thermal bath during reheating is given approximately by
\begin{align}
T_{\text{max}} \simeq \Trh \times
\begin{cases}
\left(\dfrac{a_{\text{rh}}}{a_I}\right)^{\frac{3(n - 1)}{2(n + 2)}} & \text{fermionic reheating}, \\[10pt]
\left(\dfrac{a_{\text{rh}}}{a_I}\right)^{\frac{3}{2(n + 2)}} & \text{bosonic reheating},
\end{cases}
\end{align}
which, for $a_{\text{rh}} \gg a_I$, can significantly exceed the reheating temperature $T_{\text{rh}}$, potentially by several orders of magnitude.

From \Fig\ref{fig:Trh-Tmax}, it is evident that for fermionic reheating, $\Tmax$ is above $\mathcal{O}(10^4)$ TeV for any $n$.  However, due to the constraint $\trh \ll \Tmax < \lNP$, fermionic reheating corresponds to new physics scale much above the scale which terrestrial experiments like colliders can probe. Bosonic reheating, on the other hand, allows lower $\Tmax$ and hence lower $\lNP$ with $n=4$ and 6. Therefore, in the remainder of this draft, we focus exclusively on the bosonic reheating scenario in our calculation of the DM relic density. Furthermore, we restrict most of our analysis to the low-temperature reheating regime ($\lesssim 10~\gev$), as higher temperatures would violate the validity of the effective theory, specifically, the condition $\Tmax > \lNP$.


It should be noted that preheating effects can play an important role while the Universe makes transition into radiation dominated phase after inflation \cite{Amin:2010dc,Garcia:2023eol,Garcia:2023dyf}. However, as pointed out in \cite{Kofman:1997yn,Kofman:1985aw}, perturbative decay of inflaton can still dominate the final stages of reheating, ensuring the complete transition from the inflationary to radiation phase. In fact, for $n=2$ perturbative reheating is valid for $\mu_\varphi/m_\phi \lesssim 10^{-5}$ (bosonic reheating) or $y_\psi\lesssim 10^{-5}$ (fermionic reheating) \cite{Drewes:2017fmn,Drewes:2019rxn}. As shown in \Fig\ref{fig:Trh-Tmax}, our choice of parameters agrees with these limits for $n=2$. However, for $n>2$, the non-perturbative preheating effects can become important even for the same values of these parameters $\mu_\varphi, y_\psi$ which are no longer same as $\mu_{\rm eff}, y_{\rm eff}$. Such analysis is beyond the scope of the present work and is left for future studies. On the other hand, for $n \gtrsim 8$ (corresponding to $w \gtrsim 0.65$), gravitational reheating becomes notably efficient~\cite{Haque:2022kez, Clery:2022wib, Co:2022bgh, Haque:2023yra}. In fact, under certain conditions—particularly depending on the inflaton's coupling to matter—it can even dominate over perturbative reheating, as shown in Refs.~\cite{Haque:2022kez, Haque:2023yra}. In general, for $w \gtrsim 0.65$, gravitational productions alone are capable of reheating the Universe without requiring assistance from perturbative decay channels. Nevertheless, since our analysis centers on perturbative reheating, we will confine our discussion to the regime $n<8$.

\subsection{Freeze-in Production of Dark Matter}
\label{sec:dm.prod}
The reheating phase, driven by fermionic or bosonic decay channels, produces the SM radiation bath \cite{Barman:2024ujh}. During this period, DM can be produced through a UV freeze-in mechanism. The resulting DM number density, $\ndm$, is then determined by solving the Boltzmann equation
\begin{equation} 
\frac{d\ndm}{dt} + 3 \hub \ndm = \mathcal{C}_{\text{int}},
\label{eq:beq.dm}
\end{equation}
where $\mathcal{C}_{\text{int}}$ is the DM production reaction rate density, which is expressed as \cite{Elahi:2014fsa,Bernal:2019mhf,Kaneta:2019zgw,Barman:2023ktz}
\begin{equation}
\mathcal{C}_{\text{int}}=\frac{T^{k+6}}{\Lambda_{\rm NP}^{k+2}}\,. 
\label{eq:dm.rate}
\end{equation}
Here $k=2\,(d-5)$ and $d$ is the dimension of the relevant effective DM-SM operator ($d \geq 5$). We consider the above-mentioned parametrization of the DM reaction density to derive an approximate analytical expression for the DM yield. Since SM entropy is not conserved during reheating due to inflaton annihilations, it is convenient to define the comoving number density, $\Ndm \equiv n_{\rm DM} a^3$, allowing Eq\,.~\eqref{eq:beq.dm} to be recast as
\begin{equation}
\frac{d\Ndm}{da} = \frac{a^2 \mathcal{C}_{\text{int}}}{\hub},
\label{eq:beq.dm.a}
\end{equation}
which has to be numerically solved together with Eqs.~\eqref{eq:drhodt} and~\eqref{eq:rR_fer} or \eqref{eq:rR_bos}, considering the initial condition $\Ndm(a_I) = 0$. To accommodate the observed DM relic density, it is required that
\begin{equation}
Y_0\, \mdm = \Omega_{\rm DM} h^2 \, \frac{1}{s_0}\,\frac{\rho_c}{h^2} \simeq 4.3 \times 10^{-10}~\text{GeV},
\label{eq:yld.mass}
\end{equation}
where $Y_0 \equiv Y(T_0)$, $Y(T) \equiv \ndm(T)/s(T)$, $\rho_c \simeq 1.05 \times 10^{-5}\, h^2$~GeV/cm$^3$ is the critical energy density, $s_0\simeq 2.69 \times 10^3$~cm$^{-3}$ the present entropy density~\cite{ParticleDataGroup:2024cfk}, and $\Omega h^2 \simeq 0.12$ the observed DM relic abundance ~\cite{Planck:2018vyg}. 

Depending on the DM mass scale, three distinct scenarios can arise:
\begin{enumerate}[(i)]
\item If $\mdm\ll\Trh$, DM production predominantly occurs toward the end of reheating, leading to a DM number
\begin{align}
& \Ndm(\arh)\simeq\frac{2\,\sqrt{10}\,(n+2)}{\pi\,\sqrt{\gss(\Trh)}}\,\frac{M_P\,\Trh^{k+4}}{\lNP^{k+2}}\,\arh^3
\nonumber\\&\times
\begin{cases}
\frac{1}{k-n(k+2)+10}\left[1-\left(\frac{a_I}{\arh}\right)^{\frac{3(k+10-n(k+2)}{2n+4}}\right], & \text{fermionic reheating}
\\[10pt]
\frac{1}{4n-k-2}\left[1-\left(\frac{a_I}{\arh}\right)^\frac{12n-3k-6}{2n+4}\right], & \text{bosonic reheating},
\end{cases}
\label{eq:Ndm1}
\end{align}
and corresponding yield
\begin{align}
& Y(\arh)\simeq\frac{45\,\sqrt{10}\,(n+2)}{\pi^3\,\gss(\Trh)^{3/2}}\,\frac{M_P\,\Trh^{k+1}}{\lNP^{k+2}}
\nonumber\\&\times
\begin{cases}
\frac{1}{k-n\,(k+2)+10}\,\left[1-\left(a_I/\arh\right)^{\frac{3\,(k+10-n\,(k+2)}{2n+4}}\right]\,, & \text{fermionic reheating}
\\[10pt]
\frac{1}{4\,n-k-2}\,\left[1-\left(a_I/\arh\right)^\frac{12n-3k-6}{2n+4}\right]\,, & \text{bosonic reheating}\,.
\end{cases}
\label{eq:ydm1}
\end{align}
Using \Eq\eqref{eq:rpsol}, ratio of the scale factors $a_I/\arh$ can be rewritten as
\begin{align}
& a_I=\arh\times\left(\frac{\mathcal{H}(\arh)}{\mathcal{H}(a_I)}\right)^\frac{n+2}{3n}\,.    
\end{align}

\item  On the another side, for $\Trh<\mdm\ll\Tmax$, DM is produced {\it during} reheating. In this case, the DM number density can be determined by integrating \Eq\eqref{eq:beq.dm} from the $a_I$ to $\adm=a(T=\mdm)$, where
\begin{align}\label{eq:admarh}
& \adm = \arh\times\left(\frac{\Trh}{\mdm}\right)^{1/\alpha}\,,  
\end{align}
as can be calculated using \Eq\eqref{eq:Tevol}. In this case, the DM number density at $a=\adm$ is expressed as
\begin{align}\label{eq:Ndm2}
& \Ndm(\adm)\simeq\frac{2\,\sqrt{10}\,(n+2)}{\pi\,\sqrt{\gss(\mdm)}}\,\frac{M_P\,\Trh^{k+4}}{\lNP^{k+2}}\,\arh^3
\nonumber\\&\times
\begin{cases}
\left(\Trh/\mdm\right)^\frac{k-n(k+2)+10}{n-1}\frac{1}{k-n(k+2)+10}\left[1-\left(\frac{a_I}{\adm}\right)^{\frac{3(k+10-n(k+2)}{2n+4}}\right],~~~ \text{fermionic reheating}
\\[10pt]
\left(\Trh/\mdm\right)^{4n-k-2}\,\frac{1}{4\,n-k-2}\left[1-\left(\frac{a_I}{\adm}\right)^\frac{12n-3k-6}{2n+4}\right],~~~ \text{bosonic reheating}\,,
\end{cases}
\end{align}
with corresponding yield
\begin{align}\label{eq:ydm2}
& Y(\adm)\simeq\frac{45\,\sqrt{10}\,(n+2)}{\pi^3\,\gss^{3/2}(\mdm)}\,\frac{M_P\,\Trh^{k+4}}{\mdm^3\,\lNP^{k+2}}
\nonumber\\&\times
\begin{cases}
\left(\frac{\Trh}{\mdm}\right)^\frac{6+k-n\,(k+4)}{n-1}\,\frac{1}{k-n\,(k+2)+10}\,\left[1-\left(\frac{a_I}{\adm}\right)^{\frac{3\,(k+10-n\,(k+2)}{2n+4}}\right]\,, ~~~ \text{fermionic reheating}
\\[10pt]
\left(\frac{\Trh}{\mdm}\right)^{2n-k-6}\,\frac{1}{4\,n-k-2}\,\left[1-\left(\frac{a_I}{\adm}\right)^\frac{12n-3k-6}{2n+4}\right]\,, \quad \text{bosonic reheating}\,.
\end{cases}
\end{align}
 
\item If $\Lambda_{\mathrm{NP}} > m_{\mathrm{DM}} > T_{\max}$, then most of the DM production occurs around $T_{\max}$. In this work, we consider $m_{\mathrm{DM}}$ up to $\mathcal{O}(1)$ TeV and focus on the bosonic reheating scenario with $n =4~\text{and}~6$, which yields a  $T_{\max}$ up to $\mathcal{O}(10)$ TeV and $\mathcal{O}(100)$ GeV for $T_{\text{rh}}$ up to 1 GeV, as shown in \Fig\ref{fig:Trh-Tmax}. In this case, $m_{\mathrm{DM}} > T_{\max}$ implies that $m_{\mathrm{DM}}$ is also larger than $T_{\mathrm{RH}}$. For the bosonic reheating scenario with $n = 4~(6)$, the maximum temperature $T_{\max}$ takes the values $\{0.24~(8.42), 0.80~(30.81), 1.33~(50.16), 4.10~(203.01), 6.84~(360.30)\}$ TeV (GeV) corresponding to $T_{\mathrm{RH}} = \{0.01, 0.05, 0.1, 0.5, 1.0\}~\gev$, respectively. Therefore, it is evident that we do not obtain viable points for low-mass FIMPs where $m_{\mathrm{DM}} > T_{\max}$, after imposing the correct DM relic abundance constraint. However, we have shown a yield plot in $a-\Ndm$ plane at the Appendix~\ref{sec:fimp-three} where $\mdm>\Tmax$. In this regime, due to the Boltzmann suppression \cite{Chowdhury:2018tzw, Henrich:2025gsd}, $e^{-\mdm/\Tmax}$, the effective DM number density diminishes drastically.
\end{enumerate}
\begin{figure}[h]
\centering
\includegraphics[width=0.45\linewidth]{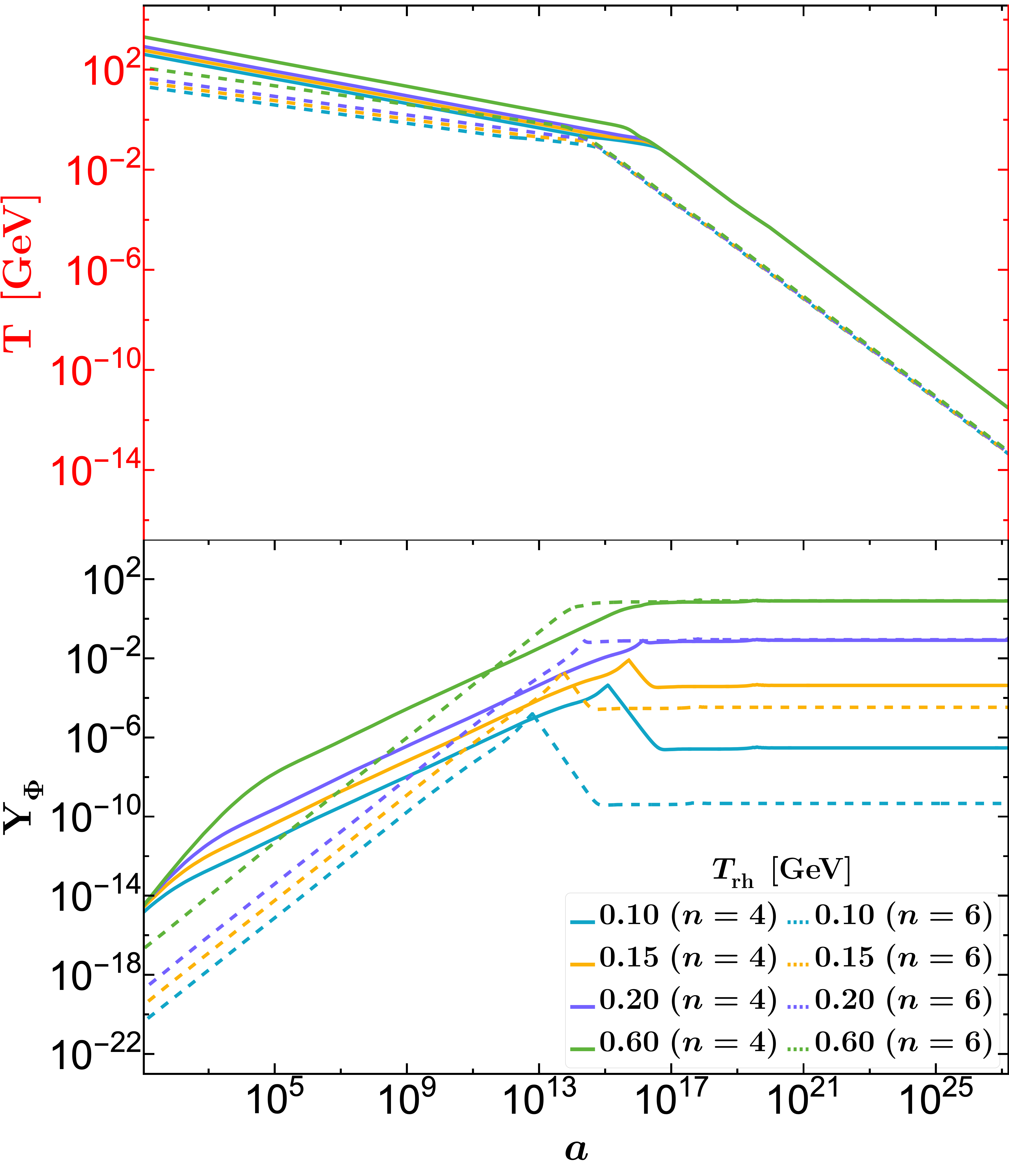}\quad
\includegraphics[width=0.45\linewidth]{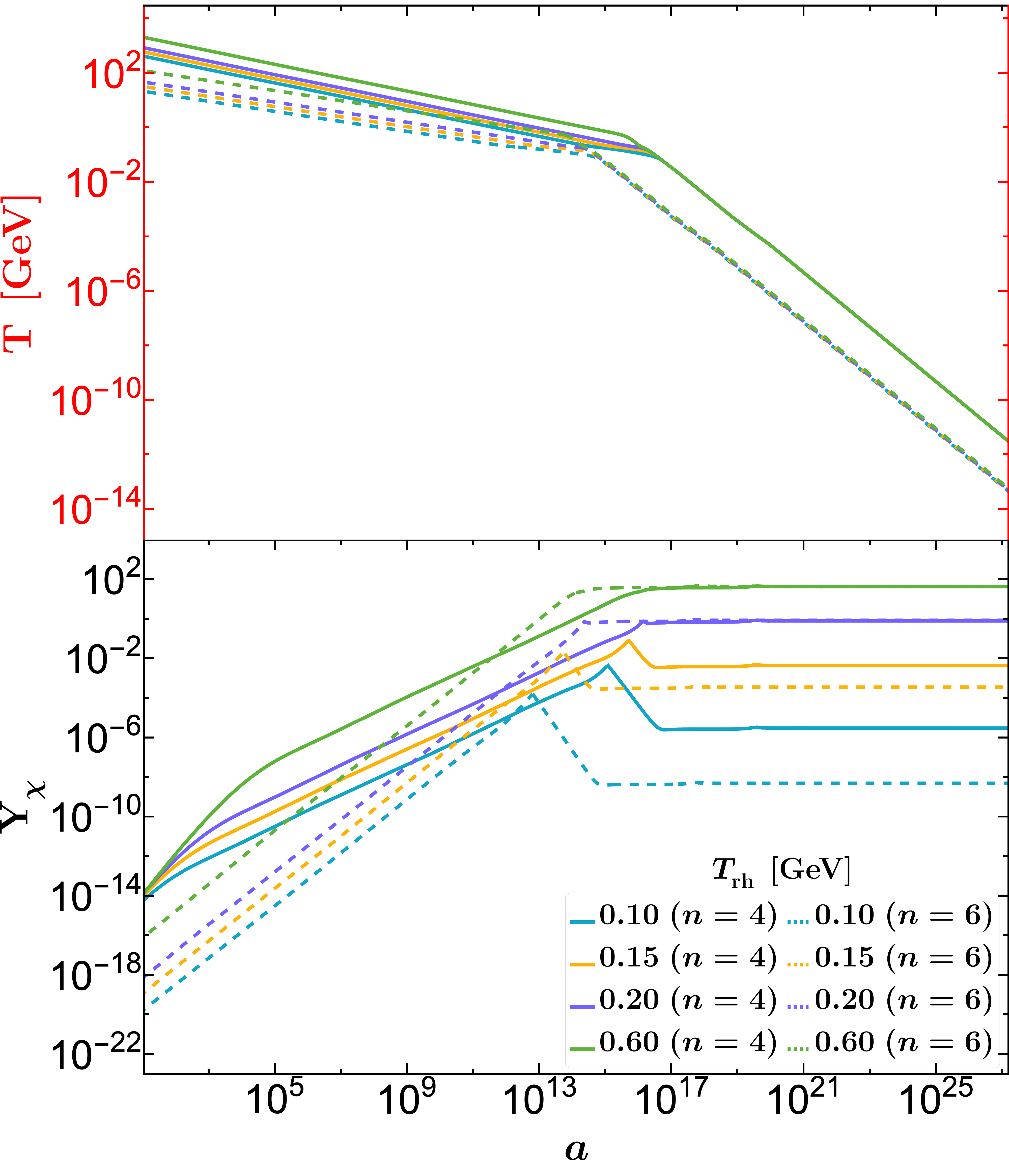}
\caption{The different colored lines correspond to various choices of reheating temperature $(\trh)$ and $n=4~(6)$ shown by solid (dashed) lines. The analysis assumes a bosonic reheating scenario, while all other parameters are fixed at $\clchi=\crchi=1.0$ for scalar DM (left) and $\clphi=\crphi=1.0$ for fermionic DM (right) with $\mdm=500~\rm MeV$ and $\lNP=2.0~\rm TeV$.}
\label{fig:yield_DM}
\end{figure}
The solution of \Eqs\eqref{eq:drhodt}, \eqref{eq:rR}, and \eqref{eq:beq.dm.a} yields the present-time DM abundance, assuming we focus solely on the bosonic reheating scenario with $n > 2$, and that the condition $\Tmax < \lNP$ is not violated. The information of the cosmological model is embedded in \Eqs\eqref{eq:drhodt} and \eqref{eq:rR}, while the DM sector appears only in \Eq\eqref{eq:beq.dm.a} through $\mathcal{C}_{\mathrm{int}}$, the DM production rate. As previously mentioned, we have adopted a monomial potential to describe the inflationary dynamics and bosonic reheating to determine the reheating temperature. Additionally, we have employed dimension-6 scalar and fermionic DM-EFT operators, see \Eqs\eqref{eq:DM-model1}-\eqref{eq:DM-model4}, to compute the DM relic abundance. Fig. \ref{fig:yield_DM} shows the evolution of comoving DM density and radiation temperature for $n=4$ as well as $n=6$, considering bosonic reheating. While the evolution remains similar for bosonic and fermionic DM, the effects of $n$ and $\Trh$ on DM relic are clearly visible. For reheat temperature close to DM mass, we see a gradual increase in DM yield before it saturates to a constant number density. For sufficiently low $\Trh$ compared to DM mass, there arises a late phase where DM abundance decreases due to entropy dilution. The kink in the temperature evolution corresponds to the transition from the stiff equation of state during reheating to the standard radiation era. Validation of non-thermal DM production for different benchmark values and evolution of DM number density are discussed in Appendix~\ref{app:non_thermal_fimp} and Appendix~\ref{sec:fimp-three}, respectively. It is worth mentioning that for DM production below the QCD confinement temperature $T_{\rm QCD} \approx 132$ MeV \cite{HotQCD:2019xnw}, the hadronic channels can also contribute, in principle, to the production of DM relic, as shown in \cite{Bhattiprolu:2022sdd} for a hadrophilic DM model with low reheating temperature. However, we restrict our analysis to leptophilic DM, where DM couples exclusively to leptons via a dimension-6 operators.  In the leptophilic case, DM-quark interactions arise solely through loop-mediated processes, which remain suppressed in both relic abundance calculations and detection prospects, compared to the leading order leptophilic interactions. The only effect of the QCD phase transition brings in our setup is the change in relativistic degrees of freedom $g_{*s}$ which has already been taken into account while solving the Boltzmann equations.

%
\section{Constraints and Detection Prospects}
\label{sec3}
\subsection{Direct search}
\label{sec:dd}
Direct search experiments look for DM scattering off a nucleon or an electron, leading to recoil of the latter at terrestrial detectors. The leptophilic nature of DM in our setup prevents tree-level DM-nucleon scattering, though it can arise at one-loop level via photon and $Z$ exchange. DM-electron scattering, on the other hand, can arise at tree level as shown on the left panel of Fig. \ref{fig:DD}. The DM-electron scattering cross-section is given by \cite{10.3389/fphy.2019.00075}
\begin{align}
\overline{\sigma}_{\rm DM~e}\equiv \dfrac{\mu_{\rm DM ~e}^2}{16\pi m_{\rm DM}^2 m_e^2}\overline{|\mathcal{M}_{\rm DM ~e}(q)|^2}\bigg|_{|{\bf q}|^2=\alpha^2m_e^2}\,.
\end{align}
In the limit of $m_e \ll m_{\text{DM}}$\footnote{In the numerical analysis, we have considered $m_e=0.51$ MeV.}, the DM-electron scattering cross-sections for scalar and fermionic DM are expressed as
\begin{align}\label{eq:dd-scalar}
\overline{\sigma}_{\Phi e}(m_e\ll \mphi)\simeq \dfrac{3m_{e}^2}{16\pi \lNP^4}\,,\\
\overline{\sigma}_{\chi e}(m_e\ll \mchi)\simeq \dfrac{3m_{e}^2}{2\pi \lNP^4}\,.
\label{eq:dd-fermion}
\end{align}
\begin{figure}[htb!]
\centering
\begin{tikzpicture}
\begin{feynman}
\vertex (a);
\vertex [above left = 1.5 cm and 1.5cm of a] (a1){\(\rm\color{black}{DM}\)};
\vertex [below left = 1.5 cm and 1.5cm of a] (a2){\(\rm\color{black}{e}\)};
\vertex [above right = 1.5 cm and 1.5cm of a] (b1){\(\rm\color{black}{DM}\)};
\vertex [below right = 1.5 cm and 1.5cm of a] (b2){\(\rm\color{black}{e}\)};
\diagram*{
(a2) -- [fermion, line width=0.35mm,  arrow size=1.0pt, style=black, edge label={\(\rm\color{black}{}\)}] (a), 
(a1) -- [plain, arrow size=0.75pt, line width=0.35mm, style=black, edge label={\(\rm\color{black}{}\)}] (a),
(b1) -- [plain,  style=black, line width=0.35mm, edge label={\(\color{black}{}\)},arrow size=0.75pt](a),
(a) -- [fermion, style=black, line width=0.35mm, edge label={\(\color{black}{}\)}, arrow size=1.0pt](b2)};
\vertex at (a) [blob, minimum size=0.7cm, fill=gray!50] {};
\end{feynman}
\end{tikzpicture} \qquad
\begin{tikzpicture}
\begin{feynman}
\vertex (a);
\vertex [above left = 1.5 cm and 1.5cm of a] (a1){\(\rm\color{black}{DM}\)};
\vertex [below left = 1.5 cm and 1.5cm of a] (a2){\(\rm\color{black}{DM}\)};
\vertex [above right = 1.5 cm and 1.5cm of a] (b1){\(\rm\color{black}{\bar{\ell}}\)};
\vertex [below right = 1.5 cm and 1.5cm of a] (b2){\(\rm\color{black}{\ell}\)};
\diagram*{
(a2) -- [plain, line width=0.35mm,  arrow size=1.0pt, style=black, edge label={\(\rm\color{black}{}\)}] (a), 
(a1) -- [plain, line width=0.35mm,  arrow size=1.0pt, style=black, edge label={\(\rm\color{black}{}\)}] (a),
(b1) -- [fermion, line width=0.35mm, style=black, edge label={\(\color{black}{}\)}, arrow size=1.0pt](a),
(a) -- [fermion, style=black, line width=0.35mm, edge label={\(\color{black}{}\)}, arrow size=1.0pt](b2)};
\vertex at (a) [blob, minimum size=0.7cm, fill=gray!50] {};
\end{feynman}
\end{tikzpicture} 
\caption{Feynman diagrams are related to the Direct (left) and Indirect (right) detection of the complex scalar ($\Phi$) and fermionic ($\chi$) DM, where $\ell$ denotes SM leptons.}
\label{fig:DD}
\end{figure}
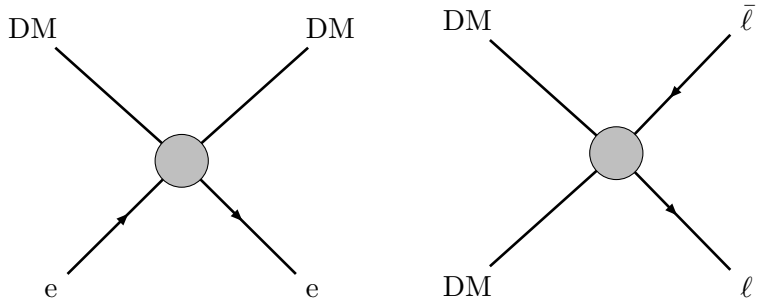
\begin{figure}[htb!]
\centering
\includegraphics[width=0.45\linewidth]{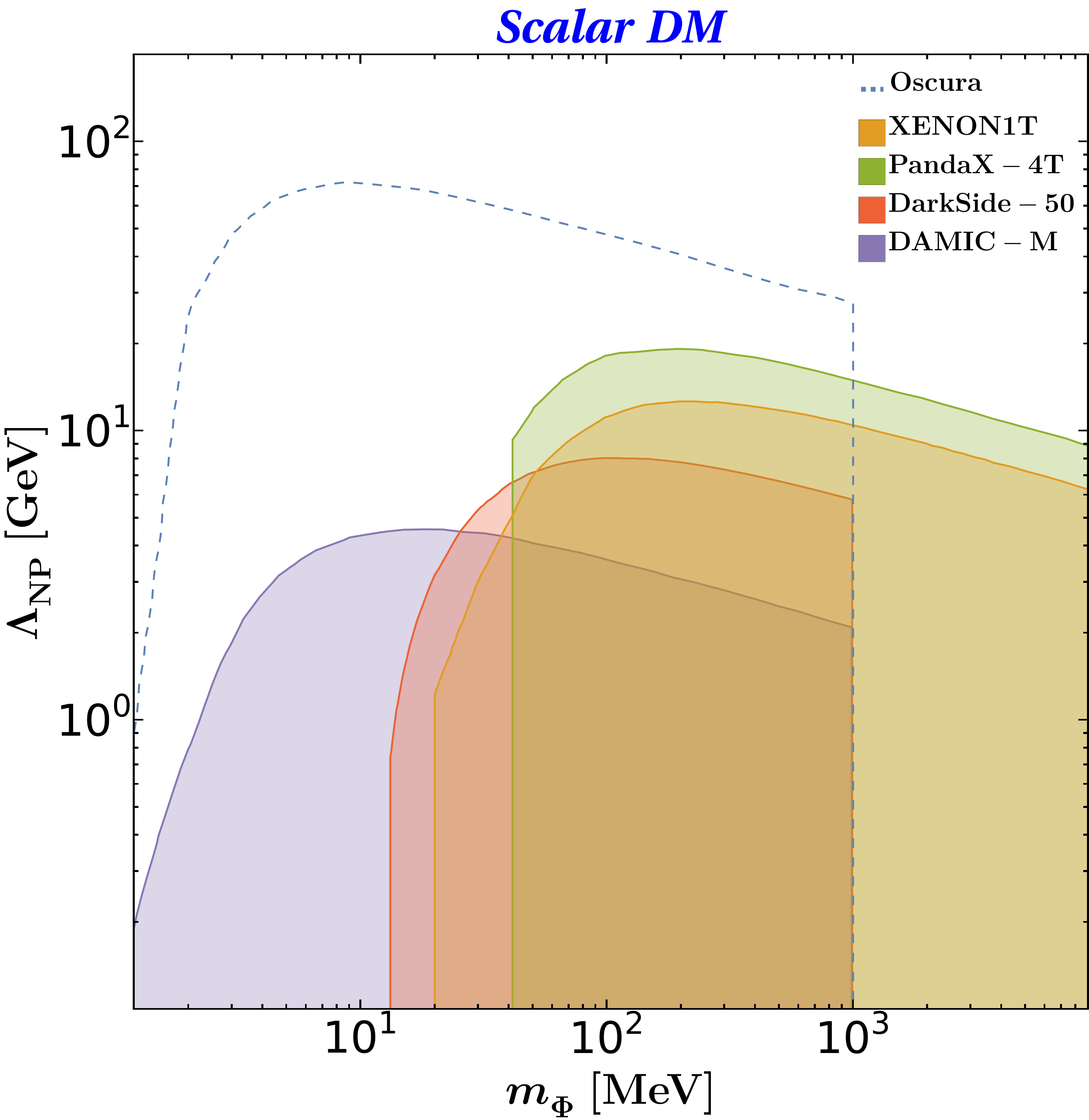}\quad
\includegraphics[width=0.45\linewidth]{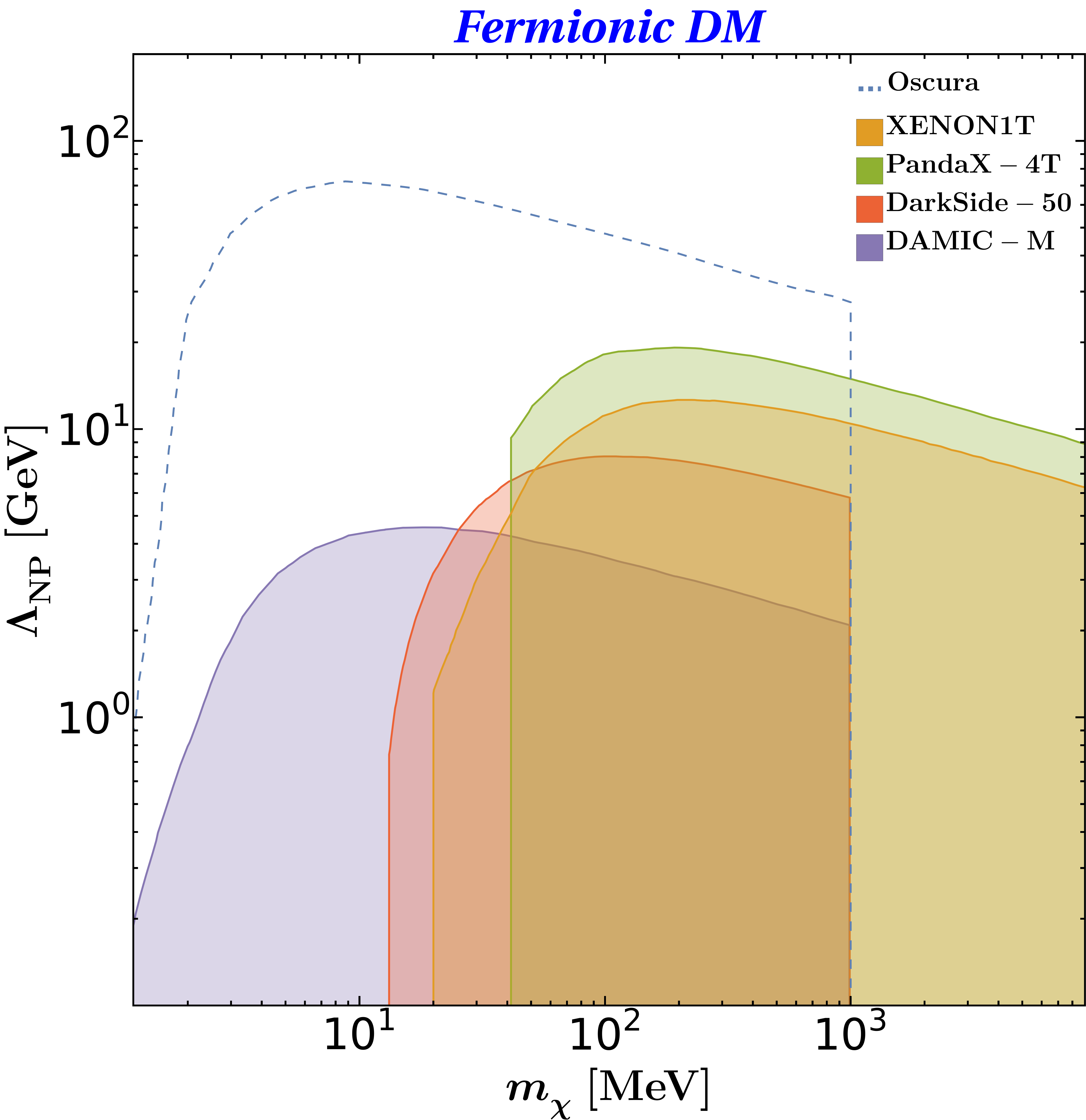}
\caption{The shaded region indicates areas excluded by the direct detection, while the thick solid (dashed) boundary shows current (future) constraints (sensitivities) on DM detection. We consider $\clchi=\crchi=1.0$ for scalar DM (left) and $\clphi=\crphi=1.0$ for fermionic DM (right).}
\label{fig:dd_summary}
\end{figure}
\noindent where we have assumed $\clchi=\clphi=1$. Since the total DM number density is the sum of the contributions from particles $(\chi, \Phi)$ and antiparticles $(\bar{\chi}, \Phi^*)$, an additional factor of one-half must be included with $\overline{\sigma}_{\rm DM~e}$. The current limits on DM-electron scattering from different experiments like XENON1T \cite{XENON:2019gfn}, DAMIC-M \cite{DAMIC-M:2025luv}, DarkSide-50 \cite{DarkSide:2022knj}, PandaX-4T \cite{PandaX:2022xqx} can constrain the parameter space in terms of DM mass and new physics scale $\lNP$ while the currently allowed parameter space can partly be probed at future experiments like Oscura \cite{Oscura:2023qik}. Fig. \ref{fig:dd_summary} shows the parameter space in $\lNP$ versus DM mass in the light of these constraints and sensitivities up to 10 GeV DM mass for demonstration purposes. Among the current experimental efforts, the PandaX-4T experiment sets the most stringent limits on the DM parameter space, constraining $\lNP$ to approximately 19 (32) GeV for scalar (fermionic) DM for a DM mass around 150 MeV. The upcoming Oscura experiment is projected to probe $\lNP$ approximately up to 73 (121) GeV scalar (fermionic) DM for a DM mass around 9 MeV. When the DM mass exceeds 10 GeV, constraints from DM–nucleon scattering become significant, as shown in Fig.~\ref{fig:summary_gw}.
%
\subsection{Indirect search}
\label{sec:id}
Although DM annihilation is negligible in the early Universe due to low reheat temperature, it can annihilate locally in the present Universe due to sizeable interaction rates with the SM bath. Such annihilations at tree level can produce lepton pairs (as shown on the right panel of Fig. \ref{fig:DD}) which further lead to diffuse photons. On the other hand, DM annihilations at one-loop can also lead to monochromatic photons. While such radiative processes are suppressed, the tree-level DM annihilation cross-sections to SM lepton pairs can be written as
\begin{align}
(\sigma v)_{\Phi\Phi^*\to e^+e^-}(m_e\ll \mphi)\simeq \dfrac{\mphi^2v^2}{6\pi \lNP^4}\left(1+\dfrac{1}{8}v^2+\mathcal{O}(v^4)\right)\,,\\
(\sigma v)_{\chi\bar{\chi}\to e^+e^-}(m_e\ll \mchi)\simeq \dfrac{\mchi^2}{\pi \lNP^4}\left(1+\dfrac{7}{24}v^2+\mathcal{O}(v^4)\right)\,,
\label{eq:ID-e}
\end{align}
\begin{align}
(\sigma v)_{\Phi\Phi^*\to \nu\bar{\nu}}\simeq \dfrac{\mphi^2v^2}{12\pi \lNP^4}\left(1+\dfrac{1}{8}v^2+\mathcal{O}(v^4)\right)\,,\\
(\sigma v)_{\chi\bar{\chi}\to \nu\bar{\nu}}\simeq \dfrac{\mchi^2}{2\pi \lNP^4}\left(1+\dfrac{7}{24}v^2+\mathcal{O}(v^4)\right)\,,
\label{eq:ID-nu}
\end{align}
where $v$ denotes the relative molar velocity of the DM particles, which is typically $\sim 10^{-3}$ in the Milky Way halo. However, near the Galactic center, it can be higher, $(2\text{--}3) \times 10^{-3}$, due to the deeper gravitational potential \cite{Gondolo:1999ef, Bringmann:2012vr}.
\begin{figure}[t]
\centering
\includegraphics[width=0.45\linewidth]{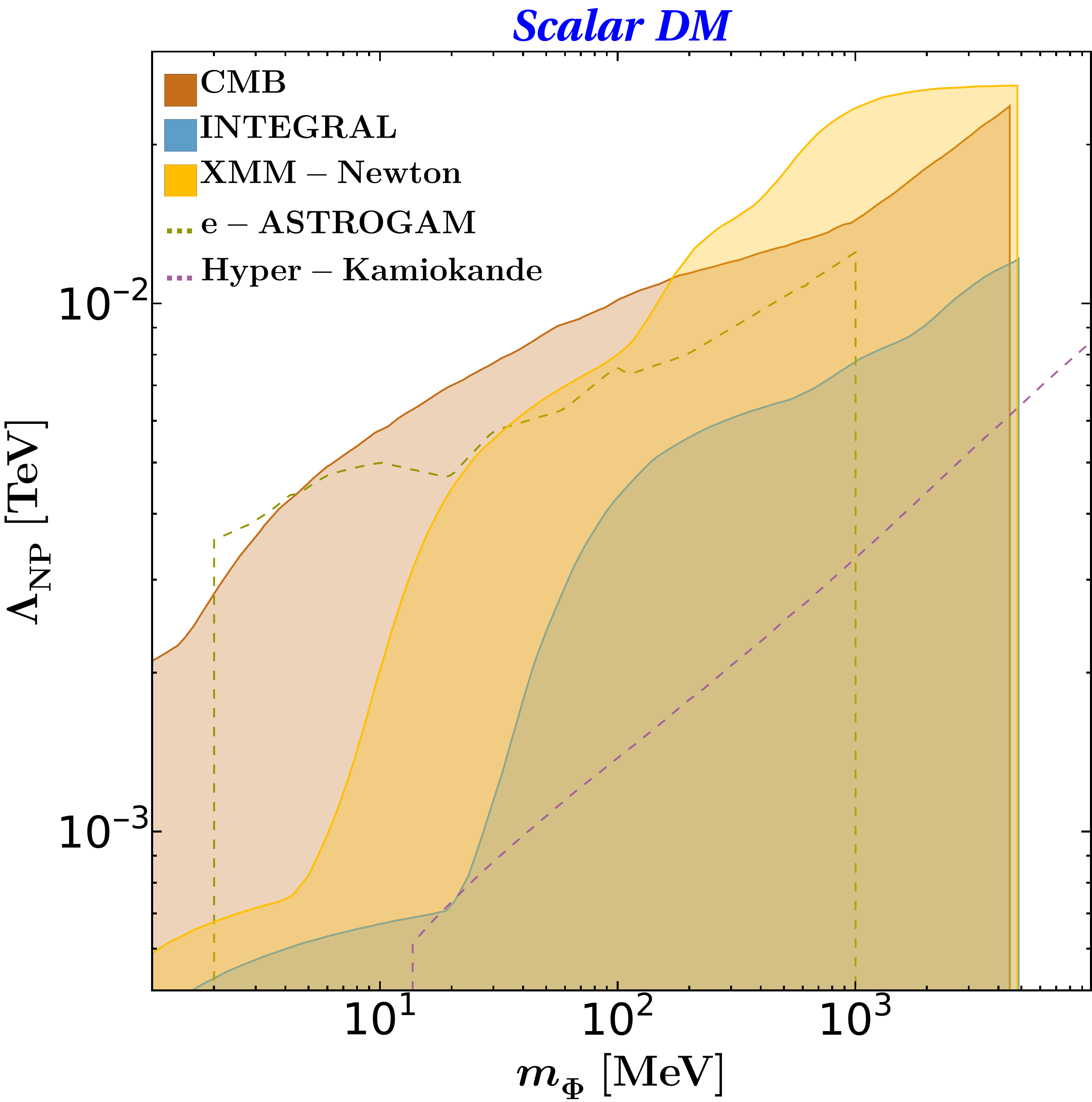}\quad
\includegraphics[width=0.45\linewidth]{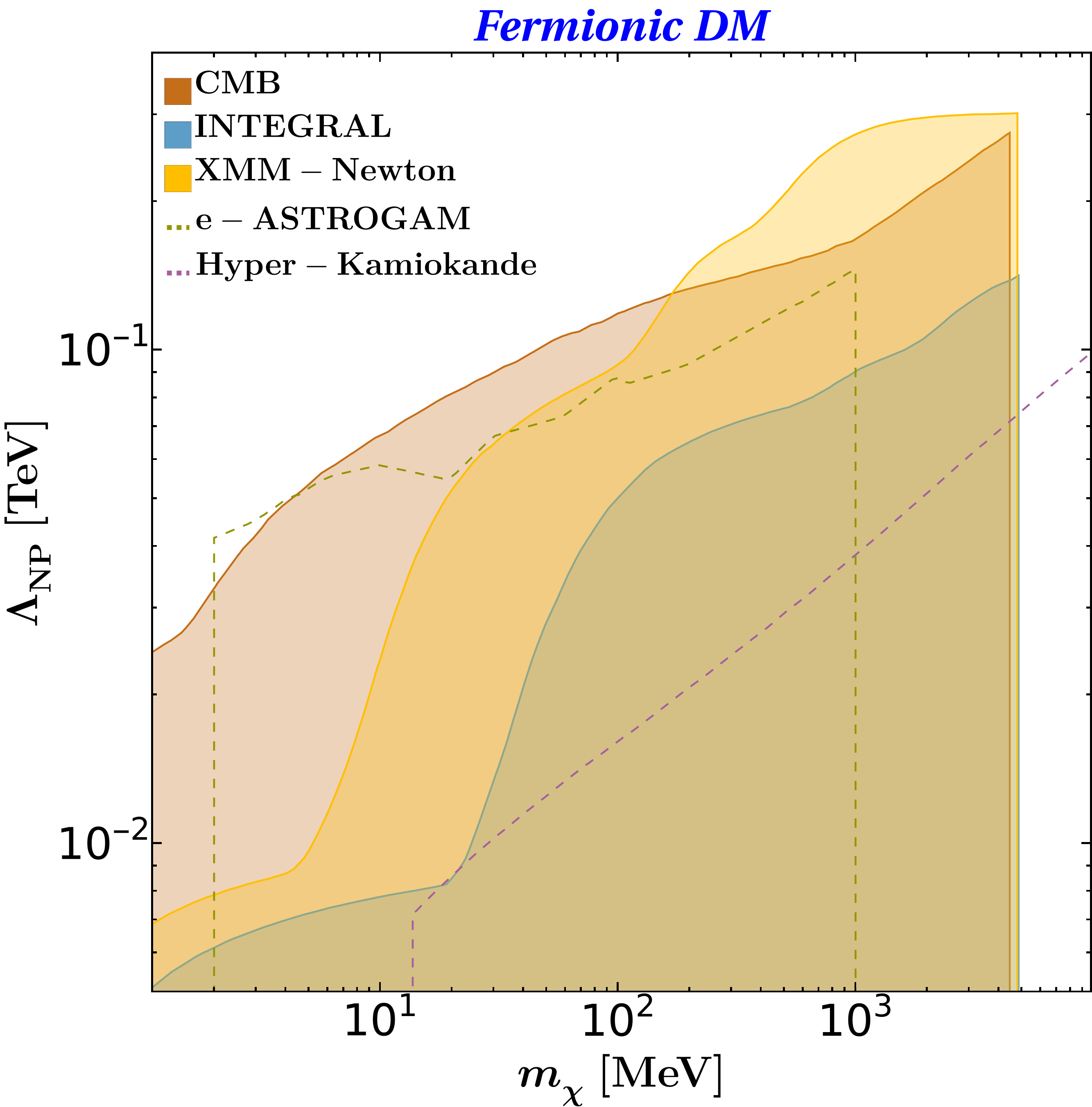}
\caption{The shaded region indicates areas excluded by the indirect detection, while the thick solid (dashed) boundary shows current (future) constraints (sensitivities) on DM annihilation. DM annihilation into a neutrino pair could also be constrained by the Hyper-Kamiokande neutrino experiment, for low $\mdm$.  For this we choose $\clchi=\crchi=1.0$ for scalar DM (left) and $\clphi=\crphi=1.0$  for fermionic DM (right).}
\label{fig:id_summary}
\end{figure}
The excess of antimatter or photons (diffuse and monochromatic) can be constrained from observations made by numerous satellites like the Fermi Large Area Telescope (Fermi-LAT) \cite{Fermi-LAT:2016afa}, INTEGRAL \cite{Cirelli:2020bpc}, XMM-Newton \cite{Cirelli:2023tnx} or future ground-based experiments like the Cherenkov Telescope Array (CTA) \cite{CTAConsortium:2012fwj}, Hyper-Kamiokande \cite{Bell:2020rkw, Arguelles:2019ouk} or space-based mission like e-ASTROGAM \cite{e-ASTROGAM:2017pxr}. Similarly, DM annihilation into neutrino pairs can be constrained from IceCube data \cite{IceCube:2023ies}. There also exist constraints from the CMB measurements on dark matter annihilation into charged final states~\cite{Madhavacheril:2013cna,Slatyer:2015jla,Planck:2018vyg}. Fig. \ref{fig:id_summary} shows a comparison of these indirect constraints and sensitivities in $\lNP-m_{\text{DM}}$ plane up to DM mass of 5 GeV for demonstration purposes. The CMB measurements place the most stringent constraints in the $\lNP\text{–}m_{\text{DM}}$ plane for DM mass below approximately 170 MeV. For masses above 170 MeV, the strongest bounds come from XMM-Newton data. For DM mass above 5 GeV, the CMB measurements again provide the leading constraints, and the corresponding constraint is shown in Fig.~\ref{fig:summary_gw}. Due to the velocity suppression in the scalar DM annihilation cross-section, the indirect search constraints remain weaker compared to the ones in the case of fermionic DM.
%
\subsection{Supernova bounds}
\label{sec:sn}
Leptophilic DM can be produced copiously from electron-positron annihilation inside a supernova if DM mass is comparable to its core temperature $T_{\rm core} \sim \mathcal{O}(30)$ MeV. Such DM can escape the supernova leading to its cooling if the mean free path of DM is of the order of supernova core size $R_{\rm core} \sim \mathcal{O}(10)$ km. Supernova cooling due to such DM escape is tightly constrained from SN1987A observations \cite{Kamiokande-II:1987idp, Bionta:1987qt}. We use the \textit{Raffelt criterion} \cite{Raffelt:1996wa} on emissivity
\begin{equation}
    \dot{\varepsilon} < 10^{19} \, {\rm erg} \, {\rm g}^{-1} \, {\rm s}^{-1},
\end{equation}
to put bounds on such energy-loss mechanism following the analysis of \cite{Dreiner:2003wh, Guha:2018mli, Magill:2018jla}. This leads to exclusion of a part of the parameter space in terms of DM mass and cutoff scale $\lNP$.

The emissivity is the energy emitted by the supernova per unit time per unit volume (due to the process $e^+ (p_1) + e^- (p_2) \rightarrow {\rm DM} (k_1) + {\rm DM} (k_2)$ given by \cite{Dreiner:2003wh}
\begin{equation}
\dot{\varepsilon} (m_{\rm DM}, T_{\rm core}, \eta) = \frac{1}{\rho_{\rm SN}} \int \frac{d^3p_1 d^3p_2}{(2\pi)^6} f_1 f_2 (E_1+E_2) v_{\rm M\o{}l} \sigma,
\end{equation}
with $\eta \equiv \mu/T_{\rm core}$ being the degeneracy of the electrons and $v_{\rm M\o{}l}$ being the relative velocity. The quantity $\rho_{\rm SN}$ denotes the supernova matter density, which is taken to be $3\times10^{14}$ g cm$^{-3}$ in the following analysis. The cross-section $\sigma$ for the process is given by
\begin{gather}
\sigma (e^+ e^- \rightarrow \chi \bar{\chi}) = \frac{A}{12 \pi \lNP^4 s}(s+2m^2_\chi) (s+2m^2_e)\,,\\ \sigma (e^+ e^- \rightarrow \Phi \Phi^\dagger) = \frac{A}{48 \pi \lNP^4 s}(s-4m^2_\Phi) (s+2m^2_e)\,,
\end{gather}
where $A = \sqrt{s-4m^2_{\rm DM}}/\sqrt{s-4m^2_e}$ and $\sqrt{s}$ is the center-of-mass (CM) energy.

The \textit{Raffelt criterion} is applicable only when the DM particles produced inside the supernova core free-stream out carrying away the energy. For a larger interaction between DM particles and electrons inside the supernova, the mean free path $\lambda_{\rm DM}$ of DM can be small, which can prevent DM to free-stream. The mean free path of DM interacting with supernova electrons can be expressed as 
\begin{eqnarray}
    \lambda_{\rm DM} = \frac{1}{n_{e}\, \sigma_{e\, \rm DM \to e\, \rm DM}},
\end{eqnarray}
where $\sigma_{e\, \rm DM \to e\, \rm DM}$ denotes the DM-electron cross-section. For DM to free-stream, we use the optical depth criterion given as 
\begin{eqnarray}
    \int_{0.9 R_{\rm core}}^{R_{\rm core}} \frac{1}{\lambda_{\rm DM}} dr\lesssim \frac{2}{3}.
\end{eqnarray}
Note that for DM mass above the average supernova core temperature, $\lambda^{-1}_{\rm DM}$ should be multiplied with the effective Boltzmann suppression factor $e^{-E_{\rm DM}/T}$ to take into effect the decreasing number density for larger masses. A more recent analysis of supernova bounds based on state-of-the-art simulations can be found in \cite{Manzari:2023gkt}.

\subsection{Collider search}
\label{sec:col}
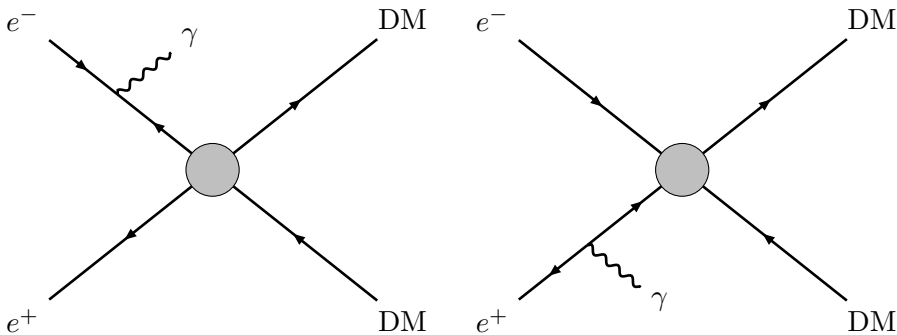
\begin{figure}[htb!]
\centering
\begin{tikzpicture}
\begin{feynman}
\vertex [dot, ultra thick] (blob) at (0,0) {};
\vertex (b)  at (-1.25,1);
\vertex (c) at (-0.3,1.75) { $\gamma$};
\vertex (g1)  at (-2.5,2) { $e^-$};
\vertex (g2) at (-2.5,-2)  { $e^+$};
\vertex (t2) at (2.5,2)  { $\text{DM}$};
\vertex (t1) at (2.5,-2)   { $\text{DM}$};
\diagram* {
(g1) -- [fermion, line width=0.35mm,  arrow size=1.0pt] (b) -- [anti fermion, line width=0.35mm,  arrow size=1.0pt] (blob) -- [fermion, line width=0.35mm,  arrow size=1.0pt] (g2),
(c) -- [photon, line width=0.35mm,  arrow size=1.0pt] (b),
(t1) -- [fermion, line width=0.35mm,  arrow size=1.0pt] (blob) -- [fermion, line width=0.35mm,  arrow size=1.0pt] (t2),
};
\vertex at (a) [blob, minimum size=0.7cm, fill=gray!50] {};
\end{feynman} 
\end{tikzpicture}\quad
\begin{tikzpicture}
\begin{feynman}
\vertex (a) at (0,0) ;
\vertex (b)  at (-1.25,-1);
\vertex (c) at (-0.3,-1.75) { $\gamma$};
\vertex (g1)  at (-2.5,2) { $e^-$};
\vertex (g2) at (-2.5,-2)  { $e^+$};
\vertex (t2) at (2.5,2)  { $\text{DM}$};
\vertex (t1) at (2.5,-2)   { $\text{DM}$};
\diagram* {
(g1) -- [fermion, line width=0.35mm,  arrow size=1.0pt]  (blob) -- [anti fermion, line width=0.35mm,  arrow size=1.0pt] (b)  -- [fermion, line width=0.35mm,  arrow size=1.0pt] (g2),
(c) -- [photon, line width=0.35mm,  arrow size=1.0pt] (b),
(t1) -- [fermion, line width=0.35mm,  arrow size=1.0pt]  (blob) -- [fermion, line width=0.35mm,  arrow size=1.0pt] (t2),
};
\vertex at (a) [blob, minimum size=0.7cm, fill=gray!50] {};
\end{feynman}
\end{tikzpicture}
\caption{Representive Feynman diagrams of mono$-\gamma$ + missing energy for DM signal at the the $e^+e^-$ colliders.}
\label{fig:monoph}
\end{figure}

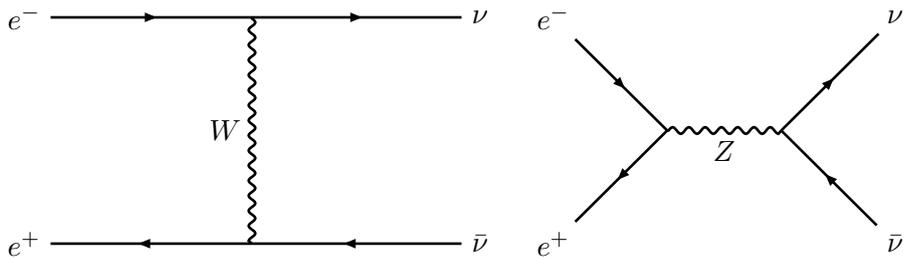
\begin{figure}[htb!]
\centering
\begin{tikzpicture}
\begin{feynman}
\vertex (a) at (-1, 1.5) {$e^-$};
\vertex (b) at (-1, -1.5) {$e^+$};
\vertex (c) at (2, 1.5);
\vertex (d) at (2, -1.5);
\vertex (e) at (5, 1.5) {$\nu$};
\vertex (f) at (5, -1.5) {$\bar{\nu}$};
\diagram*{(a) -- [fermion, line width=0.35mm,  arrow size=1.0pt] (c) -- [fermion, line width=0.35mm,  arrow size=1.0pt] (e),
(b) -- [anti fermion, line width=0.35mm,  arrow size=1.0pt] (d) -- [anti fermion, line width=0.35mm,  arrow size=1.0pt] (f),
(c) -- [boson, line width=0.35mm,  arrow size=1.0pt, edge label'=$W$] (d)};
\end{feynman}
\end{tikzpicture}\quad
\begin{tikzpicture}
\begin{feynman}
\vertex (a) at (0, +1.5) { $e^-$};
\vertex (b) at (0, -1.5) { $e^+$};
\vertex (c) at (1.5, 0);
\vertex (c1) at (3.0, 0);
\vertex (d) at (4.5, +1.5) { $\nu$};
\vertex (e) at (4.5, -1.5) { $\bar{\nu}$};
\diagram* {
(a) -- [fermion, line width=0.35mm, arrow size=1.0pt] (c) -- [fermion, line width=0.35mm,  arrow size=1.0pt] (b),
(e) -- [fermion, line width=0.35mm,  arrow size=1.0pt] (c1) -- [fermion, line width=0.35mm,  arrow size=1.0pt] (d),
(c) -- [boson,line width=0.35mm,  arrow size=1.0pt,edge label'=$Z$] (c1)
};
\end{feynman}
\end{tikzpicture}
\caption{Representative Feynman diagrams of irreducible SM backgrounds contributing to mono$-\gamma$ + missing energy for DM signal at the $e^+e^-$ colliders. For $W$ mediation (left panel), a photon can radiate from the initial state as well as from  the $W$ boson, contributing to a mono$-\gamma$ final state, whereas, for $Z$ mediation (right panel), a photon can radiate from the initial state, contributing to the given final state.}
\label{fig:sm.bkg.monoph}
\end{figure}

Lepton colliders offer a promising avenue to probe DM through missing energy and an associated visible particle that can be registered in the detector. Since DM particles evade direct detection in collider experiments by leaving no observable traces, their presence must be inferred through missing energy signatures. This energy imbalance in detector measurements provides crucial indirect evidence for DM production. The final state signal of our interest is mono$-$photon ($\gamma$) + missing energy (\Fig\ref{fig:monoph}). Two primary non-interfering irreducible SM backgrounds contribute to this final state. The first involves $W$ boson-mediated $t$-channel neutrino pair production with associated photon radiation, where the photon originates either from the initial state particles or the $W$ boson itself (left panel of \Fig\ref{fig:sm.bkg.monoph}). The second significant background stems from $Z$ boson-mediated $s$-channel neutrino pair production accompanied by initial-state photon emission (right panel of \Fig\ref{fig:sm.bkg.monoph}).

\begin{figure}[htb!]
\centering
\includegraphics[width=0.45\linewidth]{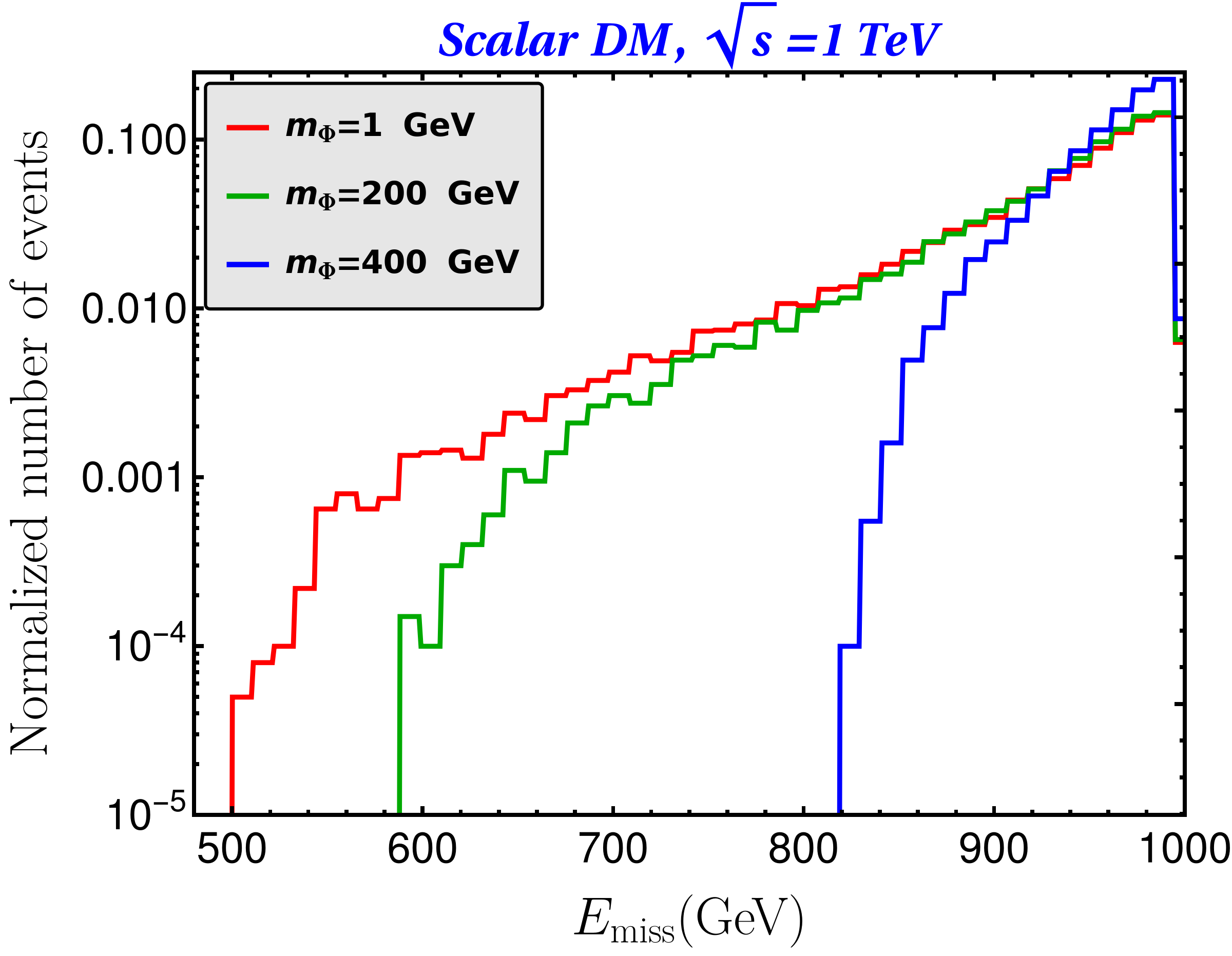}\quad
\includegraphics[width=0.45\linewidth]{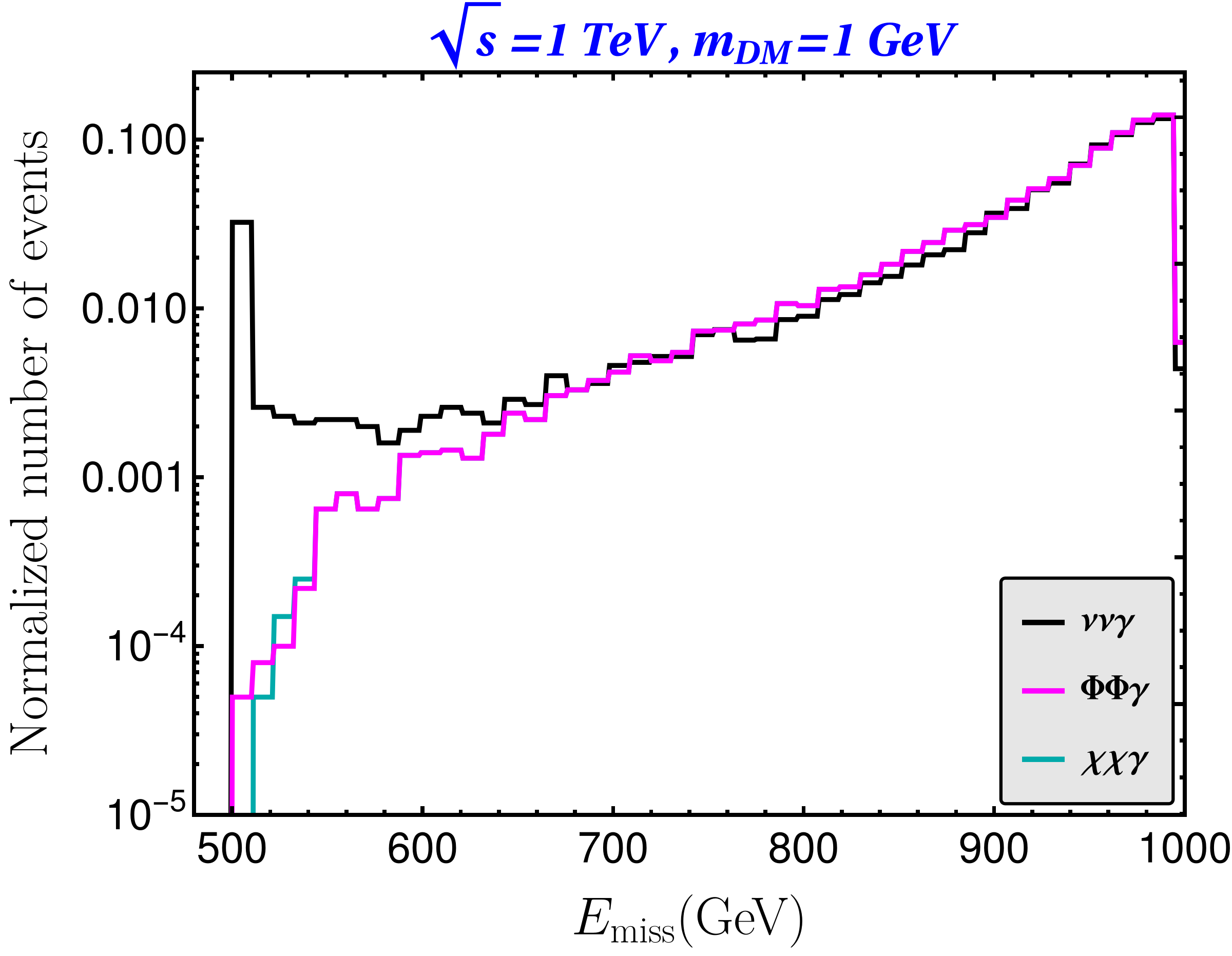}
\caption{Left panel: Normalized event distributions for the missing energy variable with different DM masses in the case of scalar DM. Right panel: Normalized signal-background event distributions for missing energy variable at $\sqrt{s}=1$ TeV ILC and DM mass of 1 GeV.}
\label{fig:event_dist}
\end{figure}

We perform a comprehensive simulation pipeline by implementing the theoretical model through {\tt FeynRules} \cite{Alloul:2013bka} and then generating signal and background events in {\tt MadGraph} \cite{Alwall:2011uj}. The generated events undergo parton showering and hadronization in {\tt Pythia} \cite{Sjostrand:2014zea} before being processed through the detector simulation in {\tt Delphes} \cite{deFavereau:2013fsa}. We impose stringent selection criteria requiring exactly one isolated photon with $p^{\gamma}_T >10$ GeV and $\eta_{\gamma}<2.5$, while explicitly vetoing events containing additional leptons or jets. For our cut-based analysis, we consider a kinematic variable, the missing energy ($E_{\text{miss}}$) and $\eta_{\gamma}$, which is defined as
\begin{itemize}
\item \textbf{Missing energy ($E_{\text{miss}}$):} The energy carried away by DM particles is known as missing energy, which is computed from the knowledge of $\sqrt{s}$ as
\beq
E_{\text{miss}}=\sqrt{s}-\sum_{i}E_{i},
\eeq
\noindent
where $i$ runs over all the detectable particles in the final state.

\end{itemize}
The normalized event distribution as a function of missing energy for various scalar DM masses is illustrated in the left panel of Fig.~\ref{fig:event_dist}. The maximum limit of the missing energy, denoted as $E^{\text{max}}_{\text{miss}}$, is determined by energy conservation and depends on the $\sqrt{s}$ and the DM mass $m_{\text{DM}}$. The expression of $E^{\text{max}}_{\text{miss}}$  is given by
\begin{equation}
E^{\text{max}}_{\text{miss}} = \frac{\sqrt{s}}{2}\left(1 + \frac{4m_{\text{DM}}^2}{s}\right).
\end{equation}
The onset of the distribution carries information about the DM mass, providing a handle to infer its value from experimental data. The $E_{\text{miss}}$ distributions for signal and background are presented in the right panel of \Fig\ref{fig:event_dist}. The higher peak of SM distribution arises due to the $W$ mediated $t$-channel process, whereas the lower peak near 500 GeV stems from the $Z$ mediated $s$-channel process. The location of the lower peak is determined by 
\begin{equation}
E_{\text{miss}}=\frac{\sqrt{s}}{2}\left(1+\frac{m_Z^2}{s}\right),
\end{equation}
where $m_Z$ is the mass of $Z$ boson. On the other hand, the signal distribution for both scalar and fermionic DM closely mimics the SM background distribution, but without the lower peak. Therefore, a cut on $E_{\text{miss}}<520$ GeV removes the $Z$-mediated contribution from the SM background.
\begin{table}[htb!]
\centering
\begin{tabular}{|c|c|c|c|c|c|c|}
\hline
\multirow{2}{*}{Cuts} & \multicolumn{3}{c|}{$\{P_{e^{+}}, P_{e^{-}}\} = \{0 \%, 0 \%\}$} & \multicolumn{3}{c|}{$\{P_{e^{+}}, P_{e^{-}}\} = \{-20 \%, +80 \%\}$} \\ \cline{2-7}
& $\Phi\Phi^{*}\gamma$ & $\nu \overline{\nu} \gamma$ & Significance & $\Phi\Phi^{*}\gamma$ & $\nu \overline{\nu} \gamma$ & Significance \\ 
\hline
Basic cuts & 1.43 & 2470 & 2.58 & 1.68  & 462 & 6.98 \\
$E_{\text{miss}} < 520$ GeV & 1.43 & 2360 & 2.63 & 1.68 & 446 & 7.10 \\
\hline
\end{tabular}
\caption{Signal and background cross-sections in fb for mono-$\gamma$ signal at the ILC with $\sqrt{s}$ = 1 TeV and $\mathfrak{L}_{\text{int}}$ = 8 $\rm{ab^{-1}}$ for unpolarized and polarized ($\{P_{e^{+}}, P_{e^{-}}\} = \{-20 \%, +80 \%\}$) beam cobinations. Here, we take scalar DM with mass $m_{\Phi} = 1$ GeV and $\Lambda_{\rm NP} = 3$ TeV.}
\label{tab:sdm}
\end{table}

\begin{table}[htb!]
\centering
\begin{tabular}{|c|c|c|c|c|c|c|}
\hline
\multirow{2}{*}{Cuts} & \multicolumn{3}{c|}{$\{P_{e^{+}}, P_{e^{-}}\} = \{0 \%, 0 \%\}$} & \multicolumn{3}{c|}{$\{P_{e^{+}}, P_{e^{-}}\} = \{-20 \%, +80 \%\}$} \\ \cline{2-7}
& $\chi\bar{\chi}\gamma$ & $\nu \overline{\nu} \gamma$ & Significance & $\chi\bar{\chi}\gamma$ & $\nu \overline{\nu} \gamma$ & Significance \\ 
\hline
Basic cuts & 5.77 & 2470 & 23.86 & 6.71  & 462 & 27.72 \\
$E_{\text{miss}} < 520$ GeV & 5.77 & 2360 & 24.28 & 6.71 & 446 & 28.21 \\
\hline
\end{tabular}
\caption{Same as Table~\ref{tab:sdm} but for fermionic DM with $m_{\chi}=1$ GeV.}
\label{tab:fdm}
\end{table}

\begin{figure}[htb!]
\centering
\includegraphics[width=0.45\linewidth]{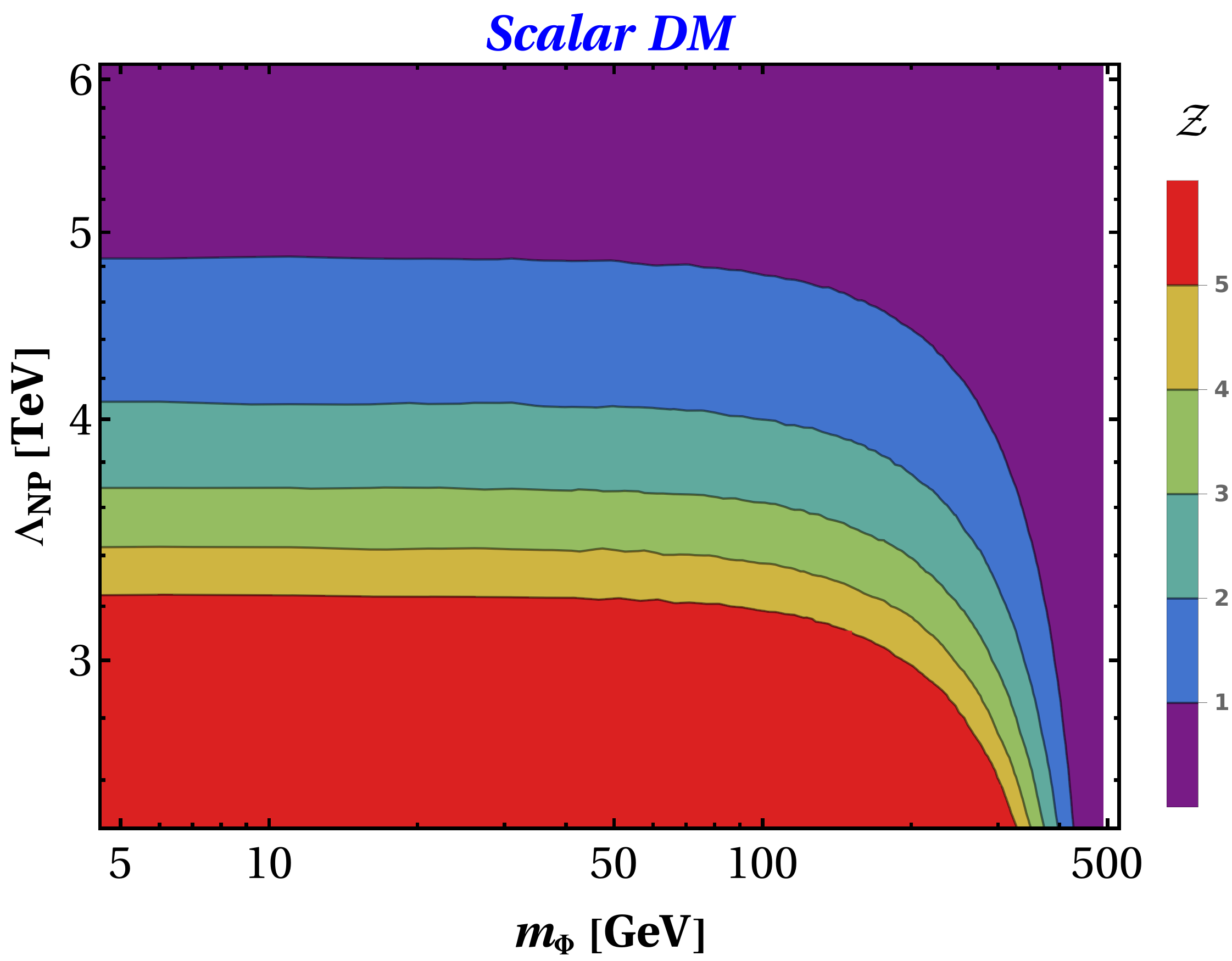}\quad
\includegraphics[width=0.45\linewidth]{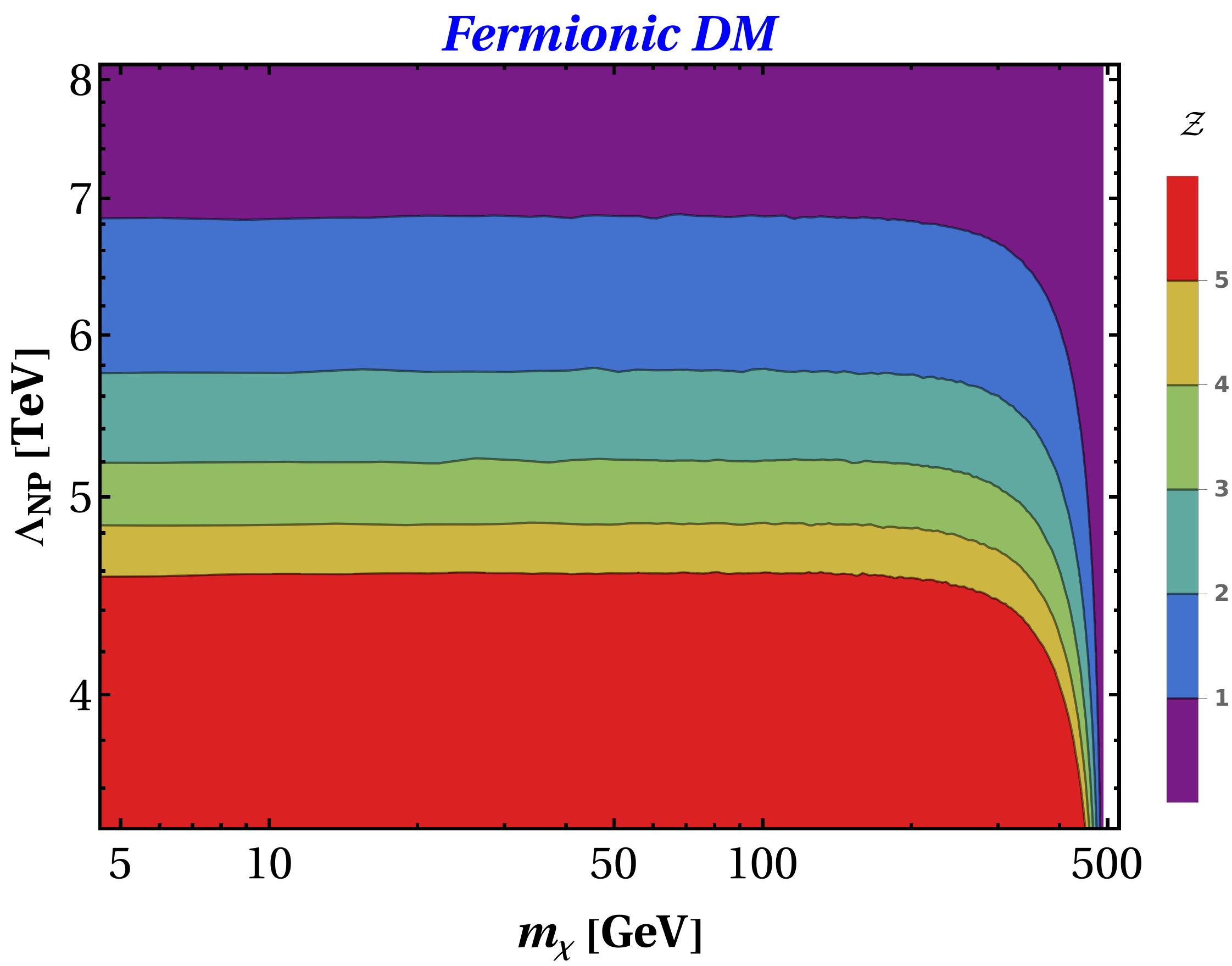}
\caption{Variation of signal significance ($\mathcal{Z})$ in $\Lambda_{\rm NP}-m_{\rm DM}$ plane at the ILC with $\sqrt{s}=1$ TeV, $\mathfrak{L}_{\rm int}=8$ ab$^{-1}$ and $\{P_{e^{+}}, P_{e^{-}}\} = \{-20 \%, +80 \%\}$ beam polarization combination. Left panel: Scalar DM, right panel: fermionic DM.}
\label{fig:significance}
\end{figure} 
Although a significant amount of SM background remains after applying kinematical cuts, initial beam polarization proves to be highly effective in further suppressing this background while enhancing the signal. Following ILC snowmass report \cite{ILCInternationalDevelopmentTeam:2022izu}, we choose $\{P_{e^+}:P_{e^-}=-20\%:+80\%\}$ which provides a six factor suppression of $\nu \bar{\nu} \gamma$ background while enhances the signal approximately by 16\% compared to the unpolarized beam. The signal significance ($\mathcal{Z}$) is defined as \cite{Cowan:2010js}
\beq
\mathcal{Z}=\sqrt{2\left[(S+B)\log\left(1+\frac{S}{B}\right)-S\right]},
\eeq
where $S$ and $B$ are the signal and background events, respectively. In the limit of $B\gg S$, $\mathcal{Z}\sim S/\sqrt{B}$ whereas for  same order of $S$ and $B$, $\mathcal{Z}\sim S/\sqrt{S+B}$. We present the variation of $\mathcal{Z}$ in $\Lambda_{\text{NP}}-m_{\text{DM}}$ plane in \Fig\ref{fig:significance} for both DM cases. As $S \propto 1/\Lambda_{\text{NP}}^4$, $\mathcal{Z}$ decreases while $\Lambda_{\text{NP}}$ increases. If we increase $m_{\text{DM}}$, $\mathcal{Z}$ decreases due to phase space suppression. Considering the ILC specification mentioned above and $\{P_{e^+}:P_{e^-}=-20\%:+80\%\}$ beam polarization, for scalar (fermionic) DM, 3$\sigma$ exclusion limit on $\lNP$ is approximately 3.7 (5.2) TeV up to a DM mass of 100 (250) GeV.

Although the above analysis is based on the ILC, we have performed a similar analysis in the context of Future Circular electron-positron Collider (FCC-ee) \cite{Agapov:2022bhm}, Compact Linear Collider (CLIC) \cite{Brunner:2022usy}, and Muon Collider ($\mu$C) \cite{Black:2022cth} as well. The maximum reach of these future lepton colliders used in this work is mentioned in Table~\ref{tab:col.reach} and 3$\sigma$ exclusion limits are presented in the summary plots (Figs.~\ref{fig:summary_scalar} and \ref{fig:summary_fermion}).
\begin{table}[hbt!]
\centering
\begin{center}
\begin{tabular}{ |c|c|c| } 
 \hline
\multirow{2}{*}{Colliders} & \multirow{2}{*}{($\sqrt{s},\mathfrak{L}_{\text{int}}$)} & Beam polarization \\ 
           & & $\{P_{e^+},P_{e^-}\}$\\
 \hline
 FCC-ee & (365 GeV, 340 $\rm{fb^{-1}}$) & $-$ \\ 
CLIC & (3 TeV, 5 $\rm{ab^{-1}}$) & $\{\pm 00\%,\pm 80\%\}$ \\ 
$\mu$C & (10 TeV, 10 $\rm{ab^{-1}}$) & $-$ \\ 
\hline
\end{tabular}
\caption{Details of the maximum reach of the proposed lepton colliders.}
\label{tab:col.reach}
\end{center}
\end{table}

\subsection{Inflationary Gravitational Wave Spectrum}
\label{sec:pgw}
The inflationary scenario of the early universe predicts the existence of primordial gravitational wave background. Upon production, the GWs interact minimally with other forms of matter and radiation and hence provide one of the cleanest probes of the early universe. The tensor metric perturbations produced during inflation are spatially stretched due to the exponential expansion of the universe making the modes superhorizon. The resulting tensor power spectrum at superhorizon scales becomes (quasi-) scale invariant. After inflation, as the horizon begins to grow faster compared to the redshifting of the length scales, the tensor modes re-enter the horizon successively. Depending on its wavelength, different modes re-enter the horizon at different periods and become sub-horizon. For the modes that re-enter the horizon during radiation domination, the resulting current inflationary GW energy density spectrum is (quasi-) scale invariant. However, the GW energy density spectrum becomes significantly titled in the frequency range corresponding to the modes that re-enter the horizon during re-heating period. The magnitude of the tilt on the energy density spectrum depends on the equation of state during re-heating period. Below, we list out the essential details of inflation GWs incorporating the re-heating period after the end of inflation and before the radiation domination \cite{Caprini:2018mtu,Figueroa:2019paj, Mishra:2021wkm, Barman:2024mqo}.

The GWs are represented by tensor-metric perturbations that satisfies the transverse and traceless conditions, $\partial_{j} h_{ij}=h^{i}_{i}=0$. The linearized Einstein equation describes the evolution of GWs and the equation, without the source term, is given as  
\begin{equation} \label{eq:gw1}
\Ddot{h}_{ij} + 3 \mathcal{H} \Dot{h}_{ij} - \frac{\nabla^2 }{a^2} h_{ij} = 0,
\end{equation}
where dots represent derivatives with respect to cosmic time. Tensor perturbations can be decomposed into Fourier- and polarization- mode as
\begin{eqnarray} \label{eq:gw2}
h_{ij}(t,\mathbf{x}) = \sum_{\lambda} \int \frac{d^{3}\mathbf{k}}{(2\pi)^3} e^{i \mathbf{k}.\mathbf{x}} \epsilon^{\lambda}_{ij}(\mathbf{k}) h^{\lambda}_{\mathbf{k}}(t).
\end{eqnarray}
The $\lambda = +, \times$ represents the two polarization states and $\epsilon^{\lambda}_{ij}(\mathbf{k})$ are the basis of polarization tensors satisfying the conditions $\epsilon^{\lambda}_{ij}(\mathbf{k}) = \epsilon^{\lambda}_{ji}(\mathbf{k})$, $\epsilon^{\lambda}_{ii}(\mathbf{k}) = k_{i}\epsilon^{\lambda}_{ij}(\mathbf{k})=0$, $\epsilon^{\lambda}_{ij}(\mathbf{k}) = {\epsilon^{\lambda}_{ij}}^{*}(\mathbf{-k})$ and $\epsilon^{\lambda}_{ij}(\mathbf{k}) {\epsilon^{\lambda'}_{ij}}^{*}(\mathbf{k}) = 2 \delta ^{\lambda \lambda'}$. With these, the equation of motion for GWs takes the form
\begin{eqnarray} \label{eq:gw3}
\Ddot{h}^{\lambda}_{k} + 3\mathcal{H} \dot{h}^{\lambda}_{k} + \frac{k^2}{a^2} h^{\lambda}_{k} = 0.
\end{eqnarray}
In the above equation, we use $h_{\mathbf{k}} = h_{k}$, with $k \equiv |\mathbf{k}|$, which is true for the isotropic background metric. 

The energy density of GWs is given by the expression 
\begin{eqnarray} \label{eq:gw5}
\rho_{\rm GW} = \frac{\langle \dot h_{ij}(t, \mathbf{x}) \dot h_{ij}(t, \mathbf{x}) \rangle}{32 \pi G} = \frac{1}{32 \pi G} \sum_{\lambda} \int \frac{d^{3}k}{(2\pi)^3} 2 |\dot{h}^{\lambda}_{k}|^2,
\end{eqnarray}
where $\langle ...\rangle$ denotes spatial average. The primordial GW spectrum is defined as the GW energy density $\rho_{\rm GW}$ per unit logarithmic comoving wavenumber interval normalized to the critical density of the universe $\rho_{\rm crit} = 3 M^2_{P} \mathcal{H}^2$,
\begin{eqnarray} \label{eq:gw6}
\Omega_{\rm GW} (t, k) \equiv \frac{1}{\rho_{\rm crit}} \frac{d \rho_{\rm GW}(t, k)}{d \, \text{ln} k}. 
\end{eqnarray}
Solving Eq. \eqref{eq:gw3}, the above expression can be written as (in terms of scale factor, $a$)
\begin{eqnarray} \label{eq:gw7}
\Omega_{\rm GW} (a, k) = \frac{k^2}{ 12 \, a^2 \mathcal{H}^2} \mathcal{P}_{T, \rm prim} \mathcal{T}(a, k). 
\end{eqnarray}
Here, $\mathcal{P}_{T, \, \rm prim}$ stands for the primordial tensor power spectrum and can be expressed in terms of primordial scalar power spectrum $\mathcal{P}_{\zeta}$ as
\begin{eqnarray} \label{eq:gw8}
\mathcal{P}_{T, \, \rm prim} = r \mathcal{P}_{\zeta} (k_{*}) \left(\frac{k}{k_{*}}\right)^{n_{T}},
\end{eqnarray}
where $k_{*}=0.05 \, \text{Mpc}^{-1}$ is the Planck pivot scale and $\mathcal{P}_{\zeta}(k_{*}) \simeq 2.1\times 10^{-9}$. The quantities $r$ and $n_{T}$ denote the tensor-to-scalar ratio and the tensor spectral index, respectively. While the current Planck results implies $r < 0.036$, for single-field inflationary models, $n_{T} \simeq -\frac{r}{8}$. As a result, we set $n_{T} = 0$ in our following analysis.   

The transfer function $\mathcal{T}(a, k)$ in Eq. \eqref{eq:gw7} connects the primordial value of the modes with the value at a later time after the modes become sub-horizon and can be expressed as 
\begin{eqnarray} \label{eq:gw9}
\mathcal{T} (a, k) = \frac{1}{2} \left(\frac{a_{\rm hc}}{a}\right)^2,
\end{eqnarray}
where $a_{\rm hc}$ is the scale factor at the time of horizon crossing for a particular mode $k$ and is quantified as $a_{\rm hc} \mathcal{H}(a_{\rm hc}) = k $. As different $k$ modes re-enter the horizon at different times, the transfer functions at today for different modes $\mathcal{T}(a_0, k)$ are different and hence bear the information of the early stages of the universe. The spectrum of GW energy density today is given as
\begin{eqnarray} \label{eq:gw10}
\Omega_{\rm GW} (k) \equiv \Omega_{\rm GW}(a_0, k) = \frac{k^2}{24 \, a^2_0 \mathcal{H}^2_0} \mathcal{P}_{T,\, \rm prim} \left(\frac{a_{\rm hc}}{a_0}\right)^2.
\end{eqnarray}

Depending on the horizon crossing of different modes $k$, the expressions for $\Omega_{\rm GW}$ and frequency $f$ can be further simplified. For inflation potential $V(\phi) \propto \phi^{n}$, the equation of state during reheating is $\omega = \frac{n-2}{n+2}$. Radiation-like epoch with $\omega = 1/3$ during reheating occurs for $n = 4$. For $n>4 \,(<4)$  implies equation of state $\omega >  1/3 \,(<1/3)$ . The expression for frequency $f$ associated with mode $k$ that crosses the horizon at $k=a_{\rm hc} \mathcal{H}_{\rm hc}$ during reheating or radiation epoch can be expressed as 
\begin{eqnarray} \label{eq:gw11}
f &\equiv& \frac{k}{2 \pi a_0} \nonumber \\
&=& \sqrt{\frac{g_{*}(T_{\rm rh})}{360}} \left(\frac{g_{*, s}(T_{0})}{g_{*, s}(T_{\rm rh})}\right)^{1/3} \frac{T_{0} T_{\rm rh}}{M_{P}} \times \begin{cases}
\left(\frac{a_{\rm rh}}{a_{\rm hc}}\right)^{\frac{2(n-1)}{n+2}} \hspace{1 cm} \text{if\,\,\,} a_{I} < a_{\rm hc} \leq a_{\rm rh}, \\ \frac{a_{\rm rh}}{a_{\rm hc}}\hspace{2.5 cm} \text{if\,\,\,} a_{\rm rh} < a_{\rm hc} \leq a_{\rm eq}.
\end{cases} 
\end{eqnarray}
Similarly, the GW energy density spectrum can be written as
\begin{eqnarray} \label{eq:gw12}
\Omega_{\rm GW}(a_{\rm hc}) &\simeq& \Omega_{\gamma, 0} \frac{\mathcal{P}_{T, \rm prim}}{24} \frac{g_{*}(T_{\rm hc})}{g^{\gamma}_{*}(T_{0})} \left(\frac{g_{* s}(T_{0})}{g_{* s}(T_{\rm hc})}\right)^{\frac{4}{3}} \nonumber \\ &\times & \begin{cases}
\frac{g_{*}(T_{\rm rh})}{g_{*}(T_{\rm hc})} \left(\frac{g_{*, s}(T_{\rm hc})}{g_{*,s}(T_{\rm rh})}\right)^{\frac{4}{3}} \left(\frac{a_{\rm rh}}{a_{\rm hc}}\right)^{\frac{2(n-4)}{n+2}} \hspace{1 cm} \text{if\,\,\,} a_{\rm I} < a_{\rm hc} \leq a_{\rm rh}, \\
1 \hspace{5.9 cm} \text{if\,\,\,} a_{\rm rh} < a_{\rm hc} \leq a_{\rm eq}.
\end{cases} 
\end{eqnarray}
Combining Eq. \eqref{eq:gw11} and Eq. \eqref{eq:gw12}, we obtain
\begin{eqnarray} \label{eq:gw13}
\Omega_{\rm GW}(f) &\simeq& \Omega_{\gamma, 0} \frac{\mathcal{P}_{T, \rm prim}}{24} \frac{g_{*}(T_{\rm hc})}{g^{\gamma}_{*}(T_{0})} \left(\frac{g_{*, s}(T_{0})}{g_{*, s}(T_{\rm hc})}\right)^{\frac{4}{3}} \nonumber \\ &\times & \begin{cases}
\frac{g_{*}(T_{\rm rh})}{g_{*}(T_{\rm hc})} \left(\frac{g_{*, s}(T_{\rm hc})}{g_{*, s} (T_{\rm rh})}\right)^{\frac{4}{3}} \left(\frac{f}{f_{\rm rh}}\right)^{\frac{n-4}{n-1}} \hspace{1 cm} \text{for\,\,\,} f_{\rm rh} < f \leq f_{\rm max}, \\
1 \hspace{5.6 cm} \text{for\,\,\,} f_{\rm eq} < f \leq f_{\rm rh}.
\end{cases} 
\end{eqnarray}
In the above equations, $\Omega_{\gamma, 0}$ and $T_{0}$ denote the fraction of photon energy density and photon temperature at the present time. The quantities $f_{\rm rh}$ and $f_{\rm max}$ correspond to the frequency of the modes that cross the horizon at the end of reheating and inflation, respectively, and are given as
\begin{eqnarray}
f_{\rm rh} &=& \sqrt{\frac{g_{*}(T_{\rm rh})}{360}} \left(\frac{g_{*, s}(T_{0})}{g_{*, s}(T_{\rm rh})}\right)^{1/3} \frac{T_{0} T_{\rm rh}}{M_{P}}, \\
f_{\rm max} &=&  \frac{H_{I}}{2\pi} \left(\frac{g_{*, s}(T_{0})}{g_{*, s}(T_{\rm rh})}\right)^{1/3} \left(\frac{\pi \, T^2_{\rm rh}}{ M_{P} H_{I}} \sqrt{\frac{g_{*s}(T_{\rm rh})}{90}}\right)^{\frac{n+2}{3n}} \frac{T_{0}}{T_{\rm rh}}. 
\end{eqnarray}

\begin{figure}[t]
\centering
\includegraphics[width=0.48\linewidth]{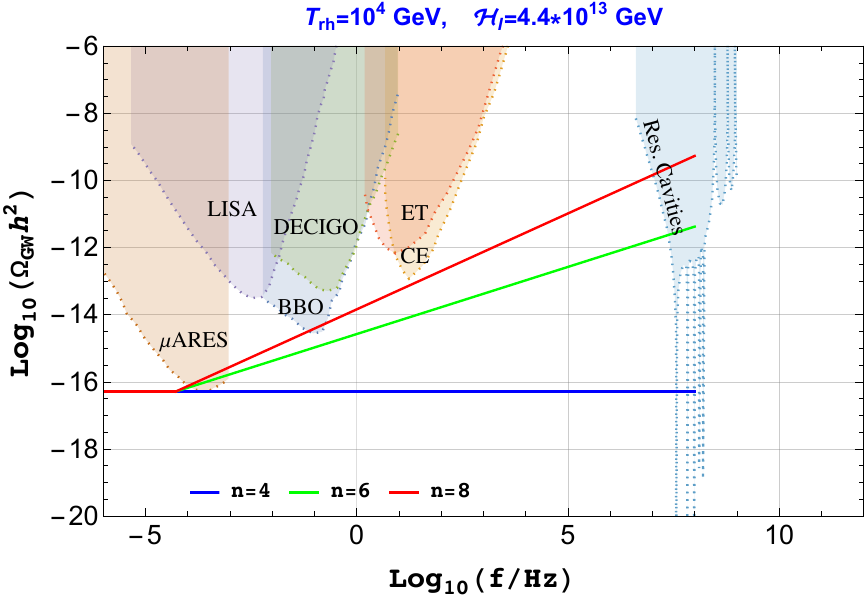}
\includegraphics[width=0.48\linewidth]{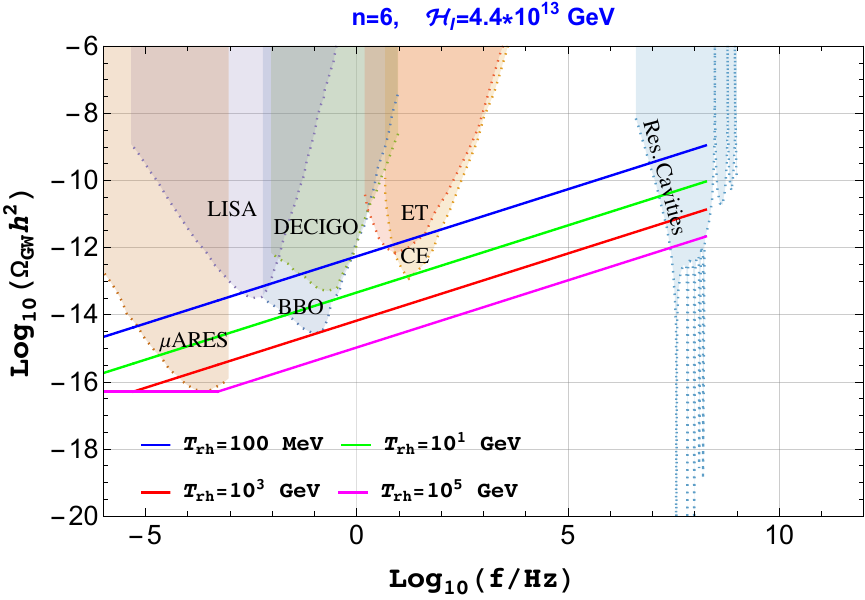}
\caption{Gravitational wave energy density spectrum as a function of frequency $f$ for different equations of state in the left panel and different reheating temperatures in the right panel. In both the panels, we consider the maximum allowed value of Hubble scale during inflation, $\mathcal{H}_{I}=4.4\times 10^{13}~\gev$.}
\label{fig:GW}
\end{figure}

Fig. \ref{fig:GW} shows the GW spectra for different values of $n$ (left panel) and $\Trh$ (right panel). The sensitivities of different future experiments such as $\mu$ARES \cite{Sesana:2019vho}, LISA \cite{amaroseoane2017laserinterferometerspaceantenna}, BBO \cite{Crowder:2005nr, Corbin:2005ny}, DECIGO \cite{Seto:2001qf, Kudoh:2005as}, ET \cite{Punturo_2010, Sathyaprakash:2012jk, ET:2019dnz} , CE \cite{Reitze:2019iox}, Resonant cavities \cite{Herman:2022fau} are shown as shaded regions.

\section{Summary and Conclusion}
\label{sec4}
\begin{figure}[htb!]
\centering
\includegraphics[width=0.405\linewidth]{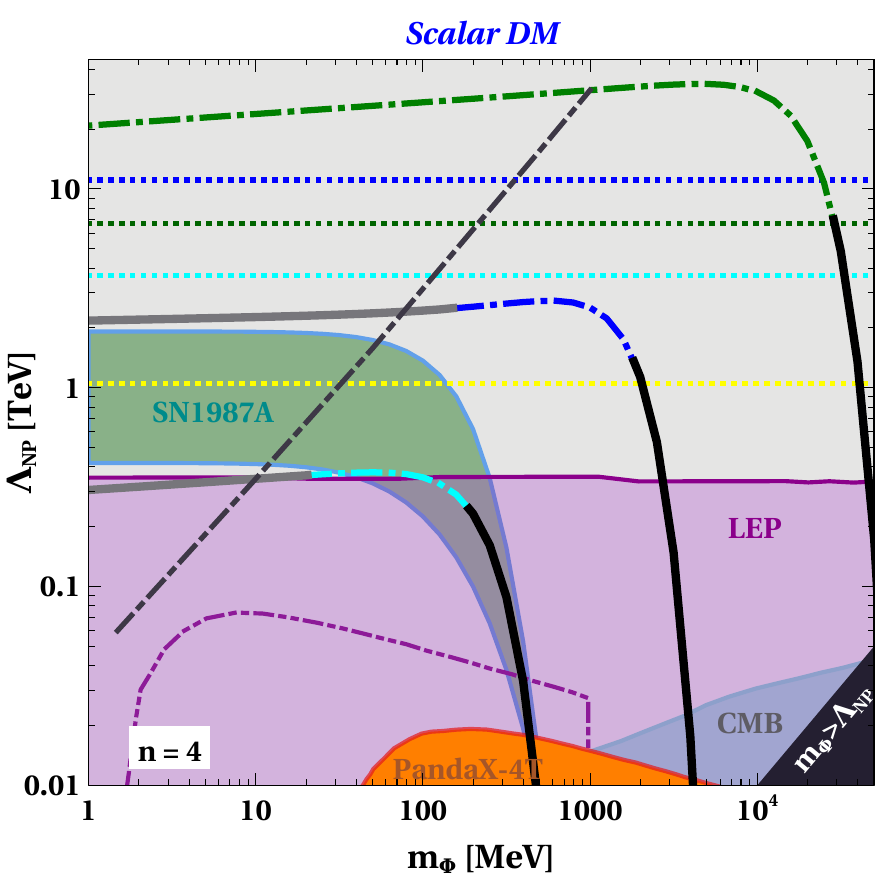}
\includegraphics[width=0.51\linewidth]{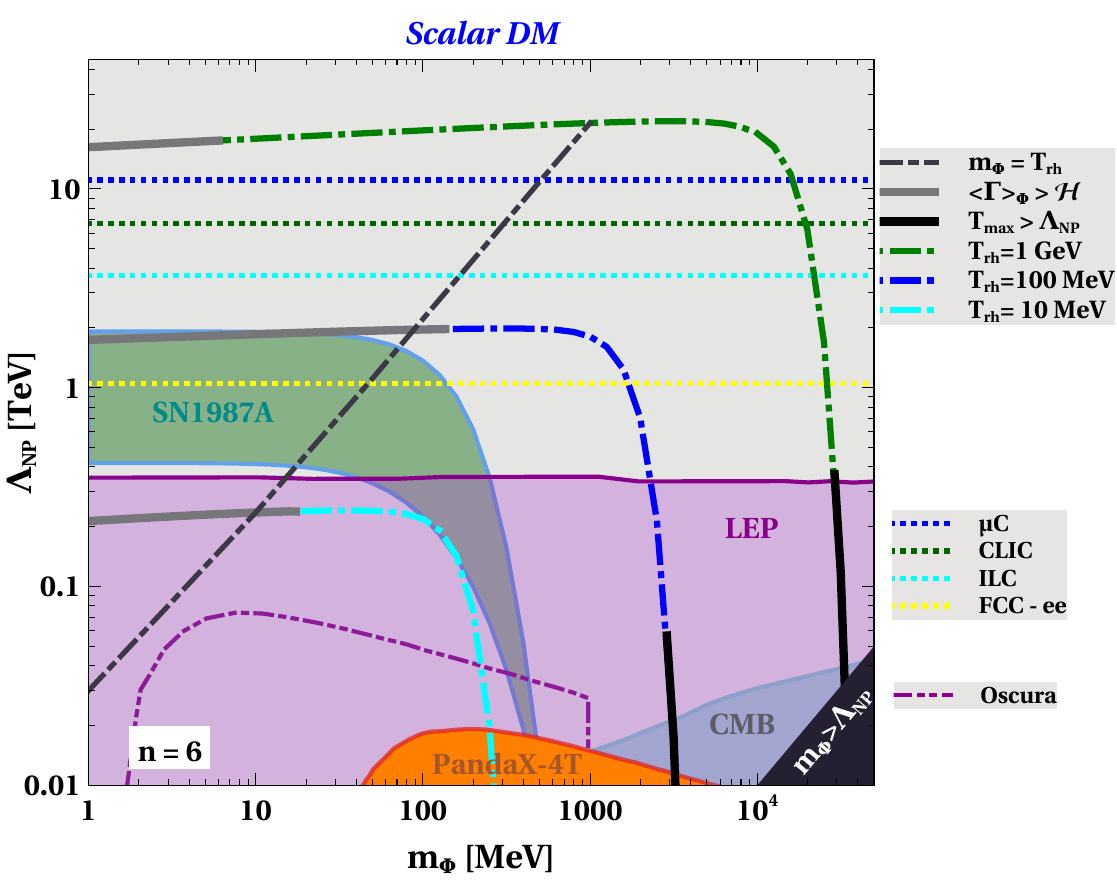}
\caption{The thick lines in the left (right) represent the relic density allowed points for scalar DM, assuming a bosonic reheating scenario with $ n=4~ ~(6)$. The black dot-dashed line represents the boundary $\mphi = \trh$, above which $\mphi < \Trh$ and below which $\mphi > \trh$. Different colors correspond to the different reheating temperatures mentioned in the figure inset. The black and gray thick shaded regions over the thick lines are disfavored by the EFT validity condition, $\Lambda_{\mathrm{NP}} > T_{\mathrm{max}}$, and by the assumption of out-of-equilibrium DM production, respectively. The shaded region indicates areas excluded by different constraints.}
\label{fig:summary_scalar}
\end{figure}

\begin{figure}[htb!]
\centering
\includegraphics[width=0.40\linewidth]{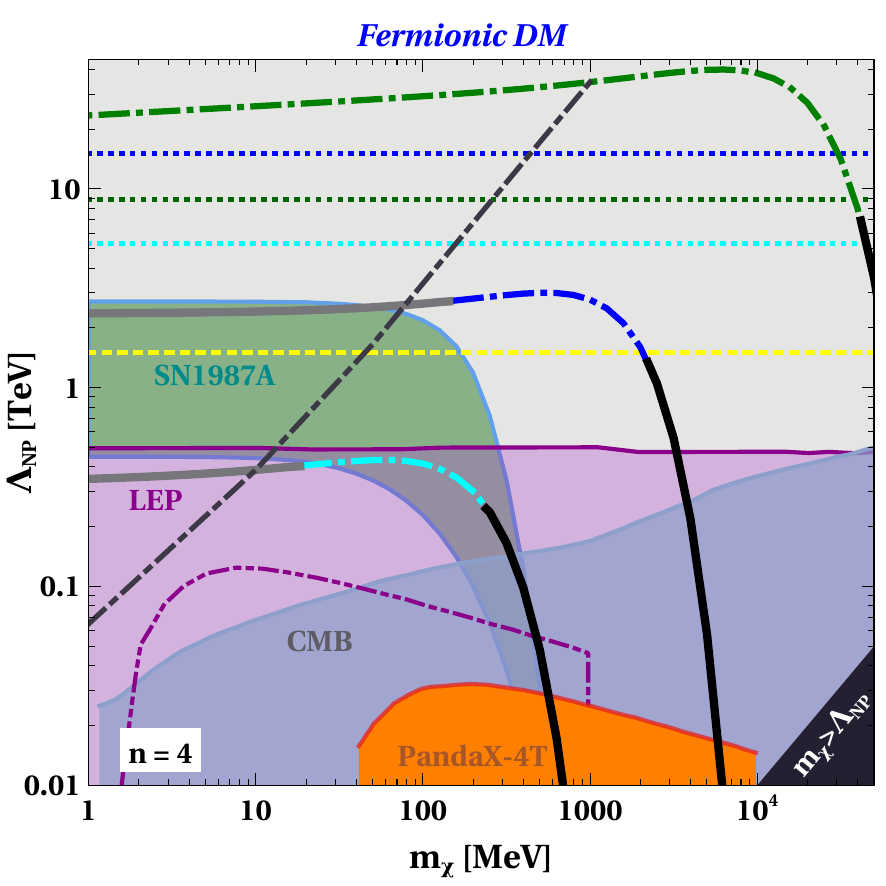}
\includegraphics[width=0.505\linewidth]{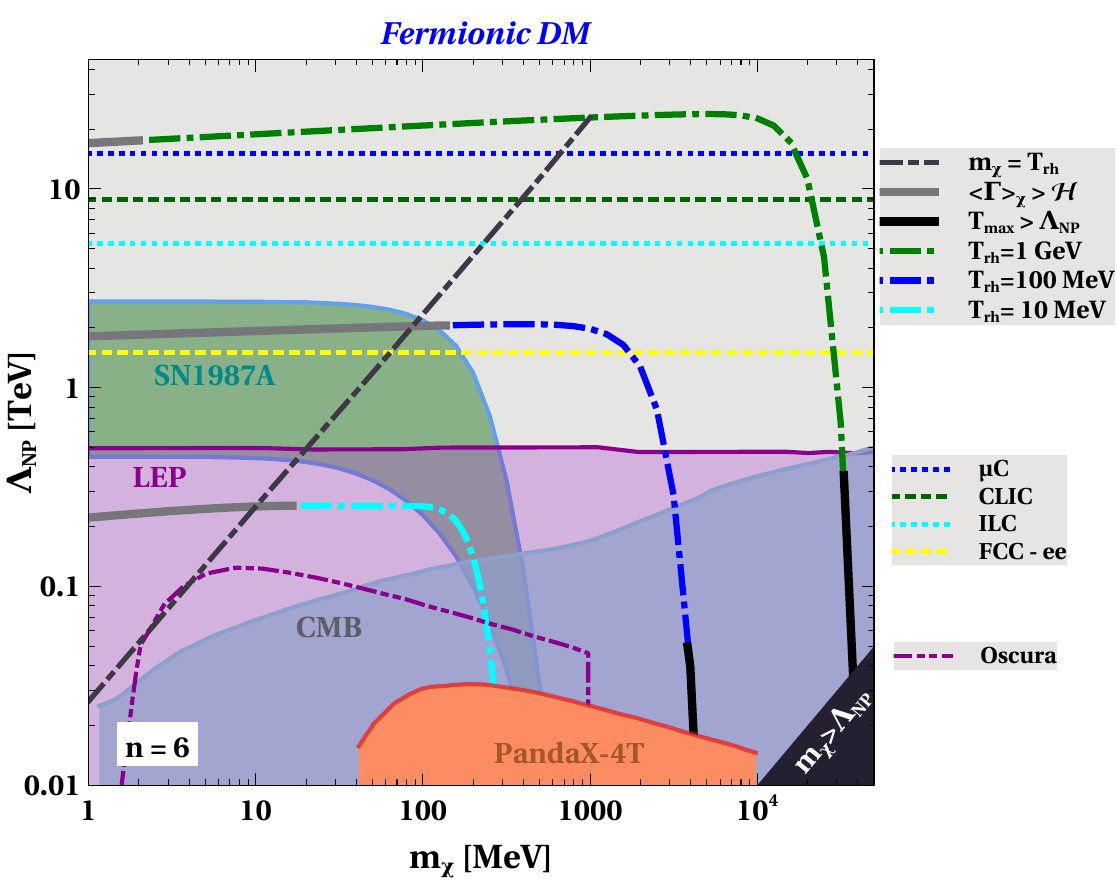}
\caption{The thick lines in the left (right) represent the relic density allowed points for fermionic DM, assuming a bosonic reheating scenario with $ n=4~ ~(6)$. The black dot-dashed line represents the boundary $\mchi = \trh$, above which $\mchi < \Trh$ and below which $\mchi > \trh$. Different colors correspond to the different reheating temperatures mentioned in the figure inset. The black and gray thick shaded regions over the thick lines are disfavored by the EFT validity condition, $\Lambda_{\mathrm{NP}} > T_{\mathrm{max}}$, and by the assumption of out-of-equilibrium DM production, respectively. The shaded region indicates areas excluded due to different constraints.}
\label{fig:summary_fermion}
\end{figure}

\begin{figure}[htb!]
\centering
\includegraphics[width=0.9\linewidth]{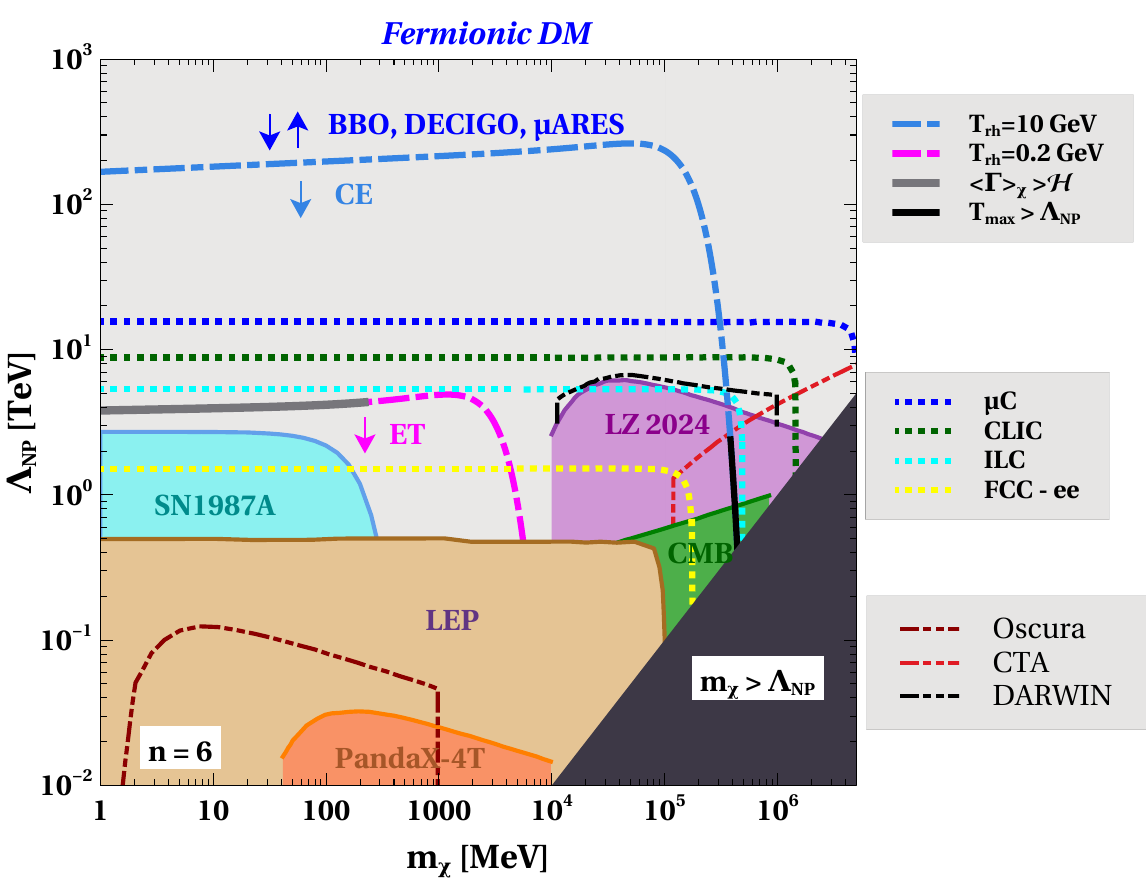}
\caption{Summary of parameter space for fermionic DM with $n=6$ including the gravitational wave sensitivities. Different dot-dashed contours correspond to the maximum $\Trh$ which the corresponding GW detector is sensitive to. The gray thick shaded part on the pink contour does not satisfy the out-of-equilibrium condition for DM. Other constraints and sensitivity contours remain same as in Fig.~\ref{fig:summary_scalar} and Fig.~\ref{fig:summary_fermion}.}
\label{fig:summary_gw}
\end{figure}

\noindent We have studied the multi-messenger discovery prospects of leptophilic DM produced non-thermally in the early Universe due to a low reheat temperature after inflation. Low $\Trh$ prevents DM thermalisation in the early Universe in spite of sizeable DM-SM interactions parametrised in terms of EFT operators of dimension-6. While sizeable DM-SM interactions keep the detection prospects promising at direct, indirect as well as collider search experiments, the non-thermal production relates DM parameter space to early Universe physics, particularly the dynamics of the reheating era after inflation. While DM-SM interactions are parametrised as dimension-6 EFT operators, the reheating era is governed by a monomial potential of the inflaton field. We study the details of DM relic generation, constraints on the parameter space and future detection prospects for different powers of the monomial potential. While the parameter space is tightly constrained from direct, indirect, and collider bounds for family-universal couplings of leptophilic DM, a large part of the currently allowed parameter space can be probed at several future experiments. Fig.~\ref{fig:summary_scalar} and Fig.~\ref{fig:summary_fermion} show the parameter space in $\lNP-m_{\rm DM}$ plane for scalar and fermionic DM, respectively, for two different choices of $n$ namely, $n=4, 6$. Here, DM-SM interactions are assumed to be family universal with vector-like DM couplings having Wilson coefficients unity. The dot-dashed contours in these figures correspond to the relic allowed parameter space for fixed reheat temperature. While the black dot-dashed line separates the regions corresponding to $m_{\rm DM} > \Trh$ and $m_{\rm DM} < \Trh$, parts of the DM relic satisfying contours (with low DM mass) are marked grey where out-of-equilibrium criteria is not satisfied. On the other hand, parts of these same contours towards high DM mass regime are marked black as those points do not satisfy the EFT validity criteria $\Tmax < \lNP$. Similarly, in the black shaded region at the right corner, the EFT is not valid since $\mdm >\lNP$. The shaded regions are ruled out by different constraints from direct, indirect search experiments, cosmology (CMB), astrophysics (supernova) as well as collider (LEP). Regions within dashed edges correspond to future sensitivities. The horizontal dashed contours of different colors correspond to $3\sigma$ exclusion limit at the future lepton colliders like muon collider (blue contour) \cite{Black:2022cth}, CLIC (green contour) \cite{Brunner:2022usy}, ILC (cyan contour) \cite{Behnke:2013xla} and FCC-ee (yellow contour) \cite{Agapov:2022bhm}. Considering family non-universal couplings will open up more allowed parameter space of these scenarios while keeping the detection prospects alive at limited number of experiments.

Interestingly, the scenarios with $n>4$ have promising detection prospects at future gravitational wave experiments. In these scenarios, the equation of state during the reheating epoch turns out to be stiffer than radiation giving a blue tilt to the primordial scale-invariant GW spectrum bringing it within reach of several experiments. Fig.~\ref{fig:summary_gw} presents the summary for fermionic DM with $n=6$ including the sensitivities of GW experiments. To show the GW sensitivities for $n=6$, we first calculate the limit on $T_{\rm rh}$ for a particular GW experiments. To calculate the limit on $T_{\rm rh}$, we use the signal-to-noise ratio (SNR) of 10, the details of which are given in Appendix \ref{sec:SNR}. Then, using the results in Fig. \ref{fig:summary_fermion}, we connect the GW sensitivities of different experiments to $\Lambda_{\rm NP}-m_{\rm DM}$ plane. Fig. \ref{fig:summary_gw} also shows the interesting complementarity among different experiments. While collider, direct, and indirect search experiments can probe upto $\lNP \lesssim 10$ TeV, some of the future GW detectors can probe much higher $\lNP$ as well. The region below the magenta dot-dashed line is within reach of future GW experiments such as ET, CE, BBO, DECIGO, and $\mu$ARES. On the other hand, the region between the blue and magenta dot-dashed lines can be probed by CE, BBO, DECIGO, and $\mu$ARES. The region above the blue dot-dashed line is accessible to BBO, DECIGO, and $\mu$ARES. Future collider sensitivities at 3$\sigma$ are shown by the dotted lines.  The current bounds from collider, direct and indirect detection experiments are shown by the shaded regions. The tan color regions show the bound from LEP experiment, while the cyan and green shaded regions indicate the indirect detection bounds from supernova and CMB, respectively. The direct detection bound from DM -electron scattering is shown by the orange region while the magenta region depicts the DM-nucleon scattering direct detection bound from LUX-ZEPLIN experiment \cite{LZ:2024zvo}. The future direct detection sensitivities from Oscura \cite{Oscura:2023qik} and DARWIN \cite{DARWIN:2024unx} are shown as brown and black dot-dot-dashed lines, respectively, while future indirect detection sensitivities for CTA experiment \cite{CTA:2020qlo} are shown by the red dot-dot-dashed line. The detection prospects of FIMP type DM scenarios at direct, indirect, and collider search experiments together with stochastic GW observations provide a multi-messenger avenue. Additionally, considering light neutrinos in the SM to be of Dirac type lead to additional DM-SM operators potentially leading to dark radiation within reach of future CMB experiments. We leave the details of such possibilities to future works.

Finally, we would like to comment on DM production through gravity-mediated processes during reheating. This kind of production is unavoidable due to universal coupling between gravity and the stress-energy tensor involving the matter particles. In our setup, gravitational production can occur directly from inflaton scattering \cite{Mambrini:2021zpp, Henrich:2024rux, Clery:2021bwz} or from scattering in the standard bath \cite{Garny:2015sjg, Tang:2017hvq, Garny:2017kha, Bernal:2018qlk, Barman:2021ugy, Barman:2021qds} both mediated by the exchange of graviton $h_{\mu\nu}$. Following the results of \cite{Henrich:2024rux, Clery:2021bwz}, we can naively conclude that such gravitational production remains sub-dominant in the parameter space shown in Fig. \ref{fig:summary_scalar}, Fig. \ref{fig:summary_fermion} and Fig. \ref{fig:summary_gw} except for scalar DM with $n=6$.
However, such conclusions also rely heavily on $\Trh$ which, if not sufficiently high, may require additional effects like inflaton fragmentation \cite{Garcia:2023dyf} for $n\geq 4$. Also note that the results presented in papers \cite{Henrich:2024rux, Clery:2021bwz} consider fermionic reheating only whereas our results are based on bosonic reheating. A detailed study of these subtle effects related to gravitational production, including specific reheating mode (bosonic and fermionic) is beyond the scope of this present work and is left for future studies.

\section*{Acknowledgement}
The work of D.B. is supported by the Science and Engineering Research Board (SERB), Government of India grants MTR/2022/000575 and CRG/2022/000603. D.B. also acknowledges the support from the Fulbright-Nehru Academic and Professional Excellence Award 2024-25. The work of N.D. is supported by the Ministry of Education, Government of India via the Prime Minister's Research Fellowship (PMRF) December 2021 scheme. N.D. thanks Rishav Roshan and Indrajit Saha for useful discussions regarding GWs. S.J. thanks Bhavya Thacker for useful discussions. 

\appendix

\section{Validation of Non-Thermal Dark Matter Production}
\label{app:non_thermal_fimp}
\begin{figure}[htb!]
\centering
\includegraphics[width=0.45\linewidth]{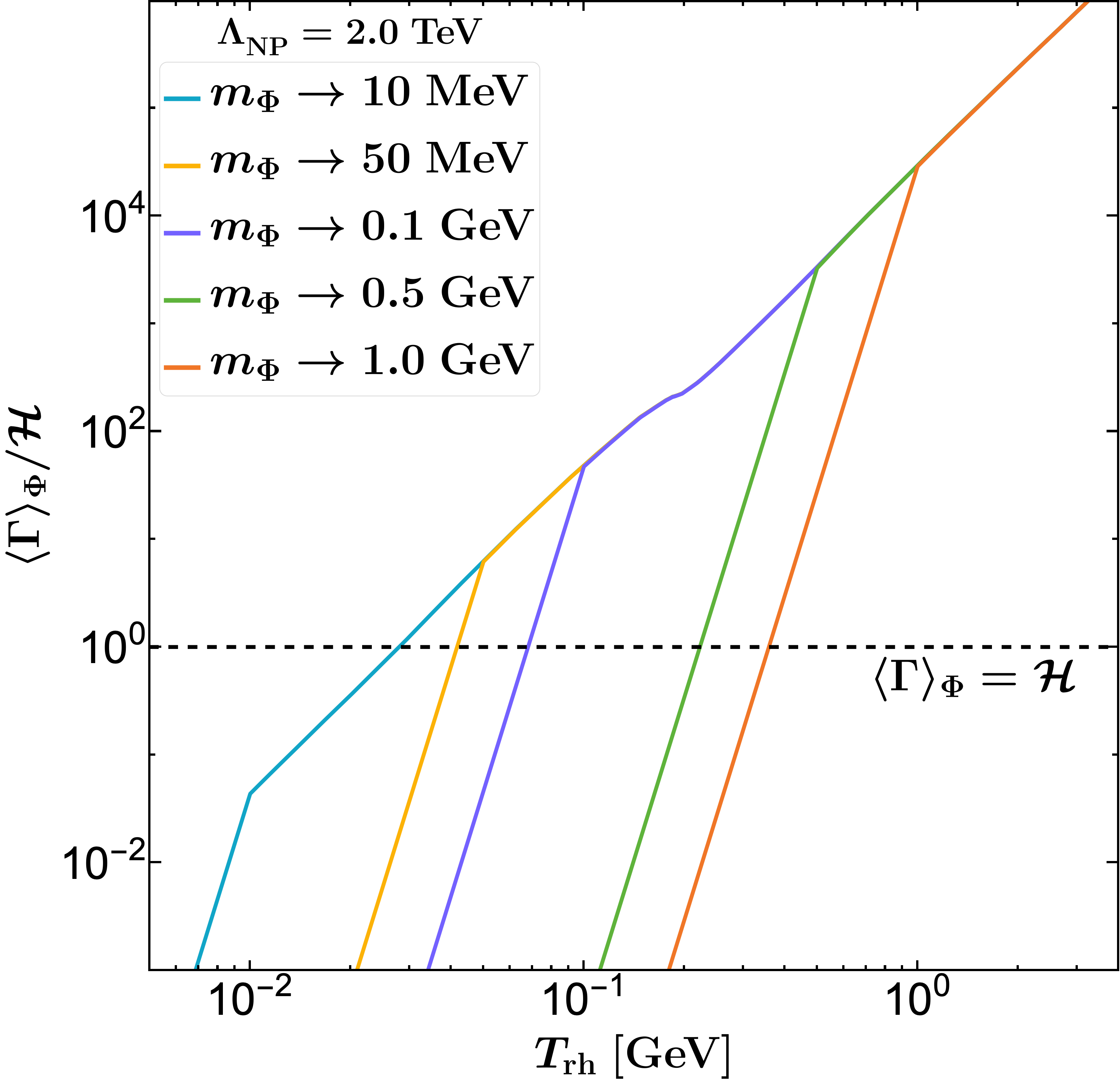}\quad
\includegraphics[width=0.45\linewidth]{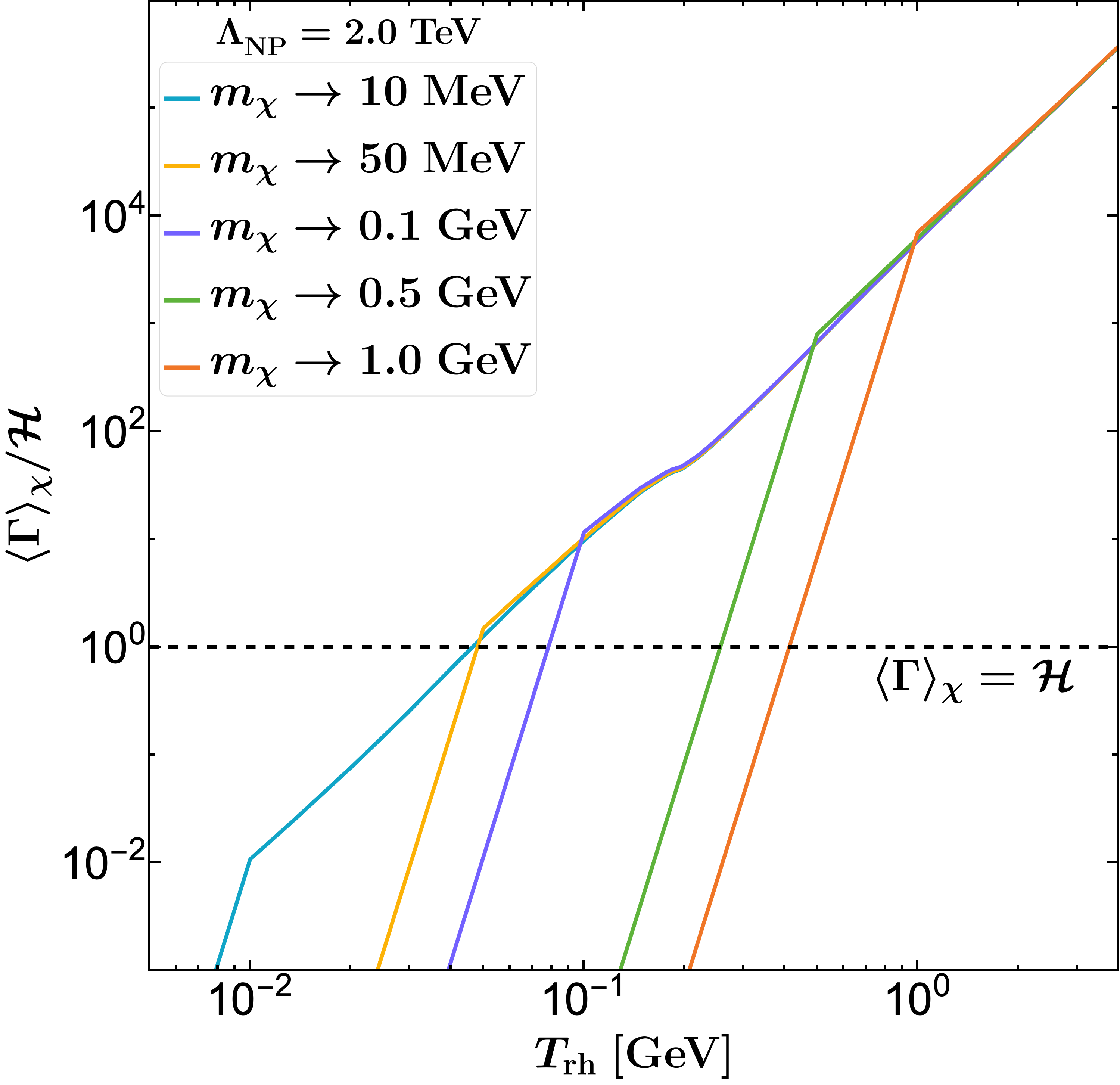}
\caption{Non-thermal production of DM is ensured when the ratio $n=6,~\langle \Gamma \rangle_{\rm DM}/\mathcal{H}$ remains less than unity during the freeze-in process. Here, we assumed all other Wilson coefficients are taken as $1.0$.}
\label{fig:gammaH}
\end{figure}

Fig. \ref{fig:gammaH} shows the rate of DM-SM interactions compared to Hubble at different reheating temperatures for $\lNP=2$ TeV and varying DM mass. In order to validate the non-thermal production mechanism of DM, we stick to a regime where $\langle\Gamma\rangle_{\rm DM} < \mathcal{H}$.
\section{FIMP Production in three different regimes}
\label{sec:fimp-three}
\begin{figure}[htb!]
\centering
\includegraphics[width=0.45\linewidth]{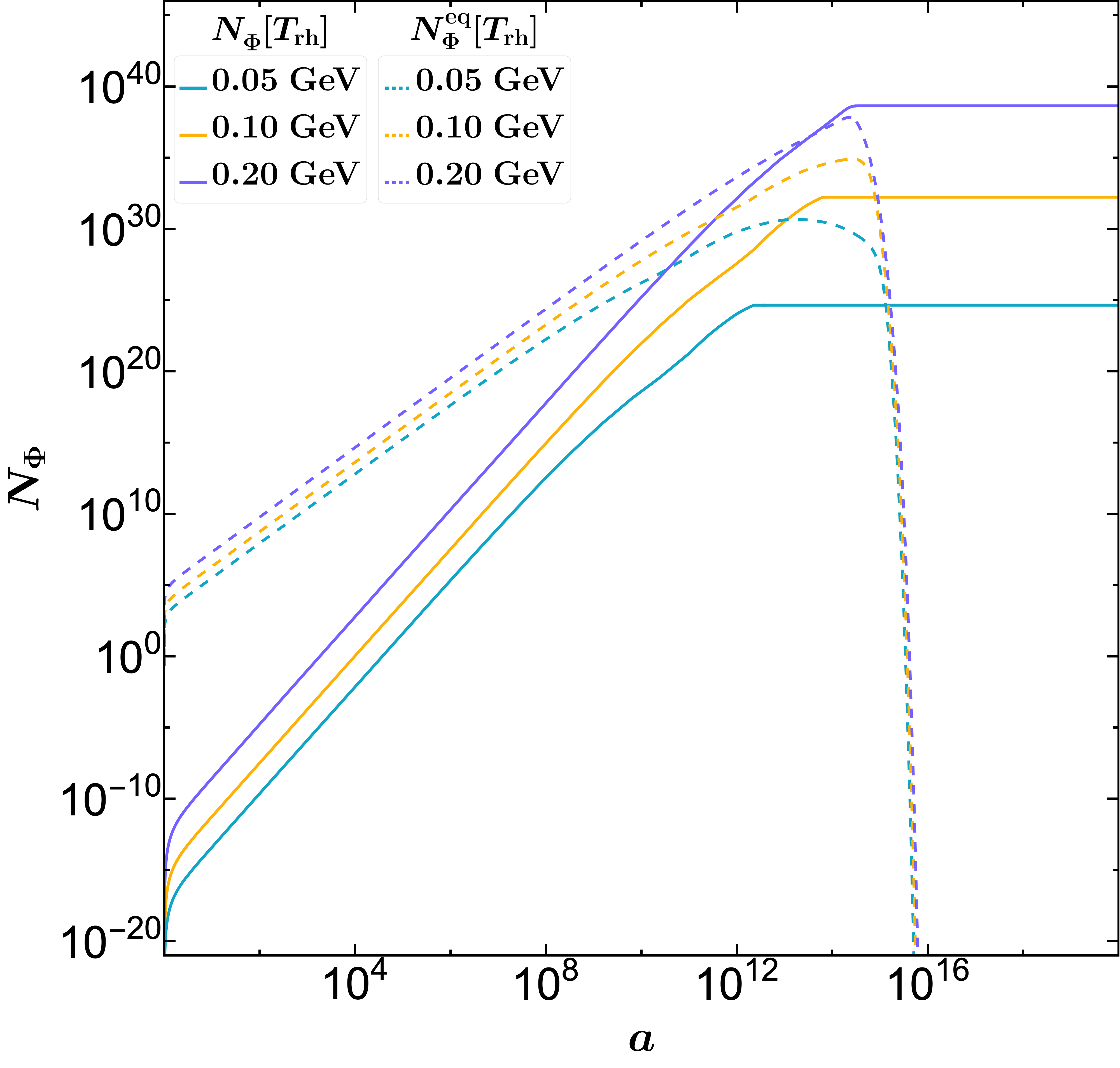}\quad
\includegraphics[width=0.45\linewidth]{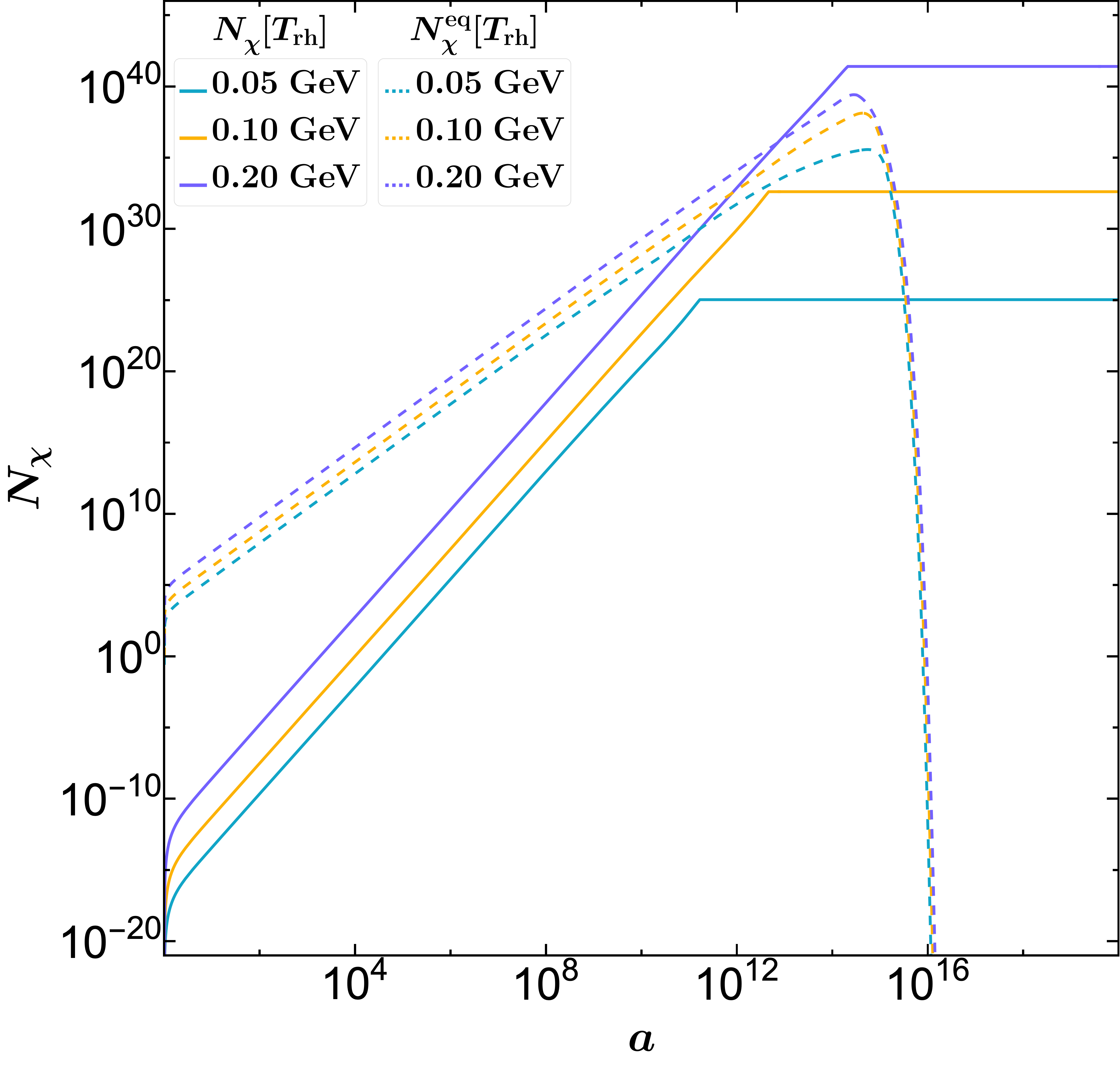}
\caption{Figures are representing the evaluation of co-moving DM number density ($N_{\rm DM}=\ndm a^3$) with the scale factor ($a$) while the dashed lines represent the co-moving equilibrium number density and $n=6$.
Here, we assumed all other Wilson coefficients are unity. Here we have fixed the parameters in the left (right) plot as: $\lNP=1450~(2074)~\gev$, $\mdm=1.42~(0.62)~\gev$. The yellow color line produces the correct observed relic density.}
\label{fig:gammaH-N}
\end{figure}

\begin{figure}[htb!]
\centering
\includegraphics[width=0.45\linewidth]{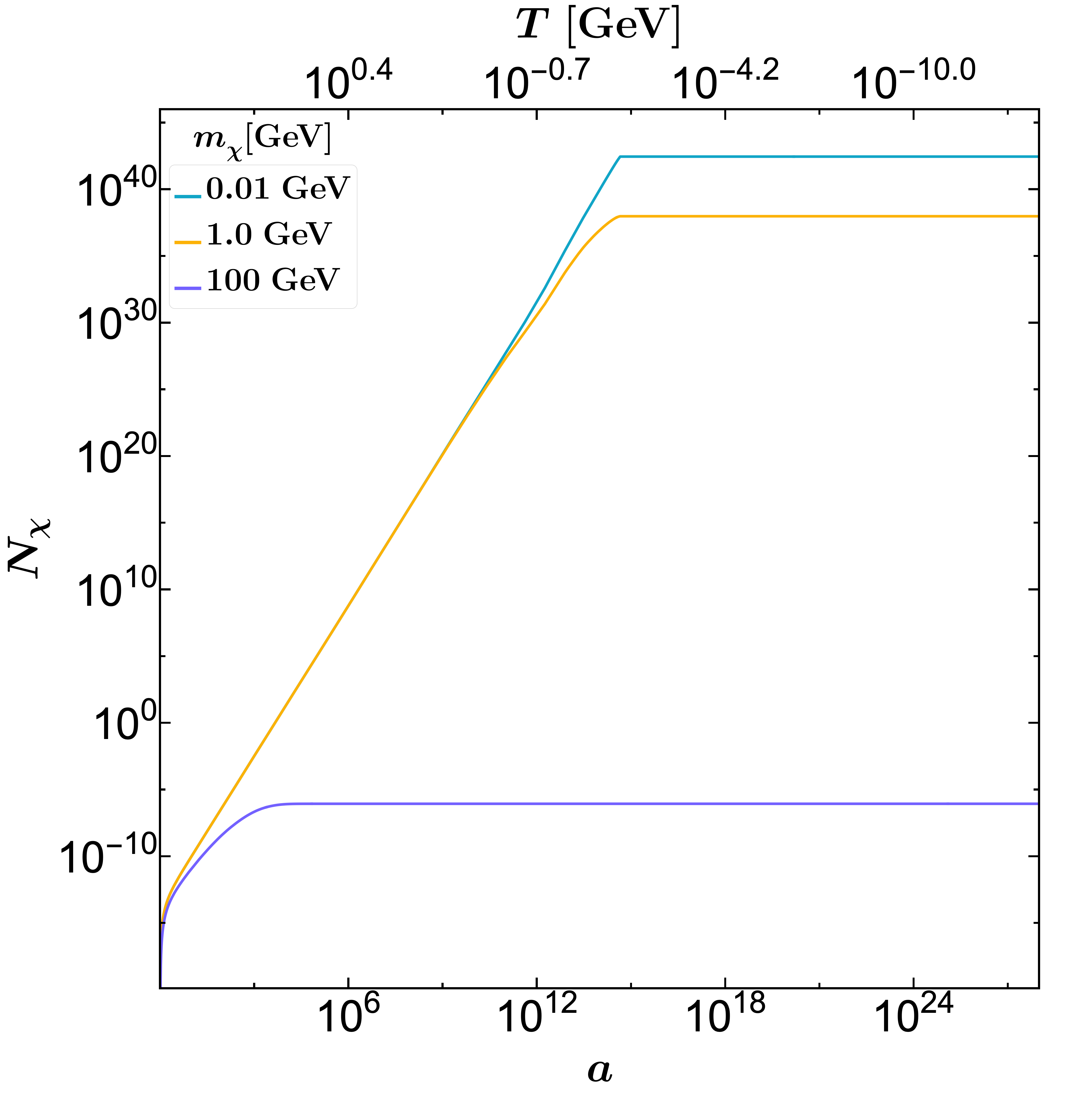}
\includegraphics[width=0.45\linewidth]{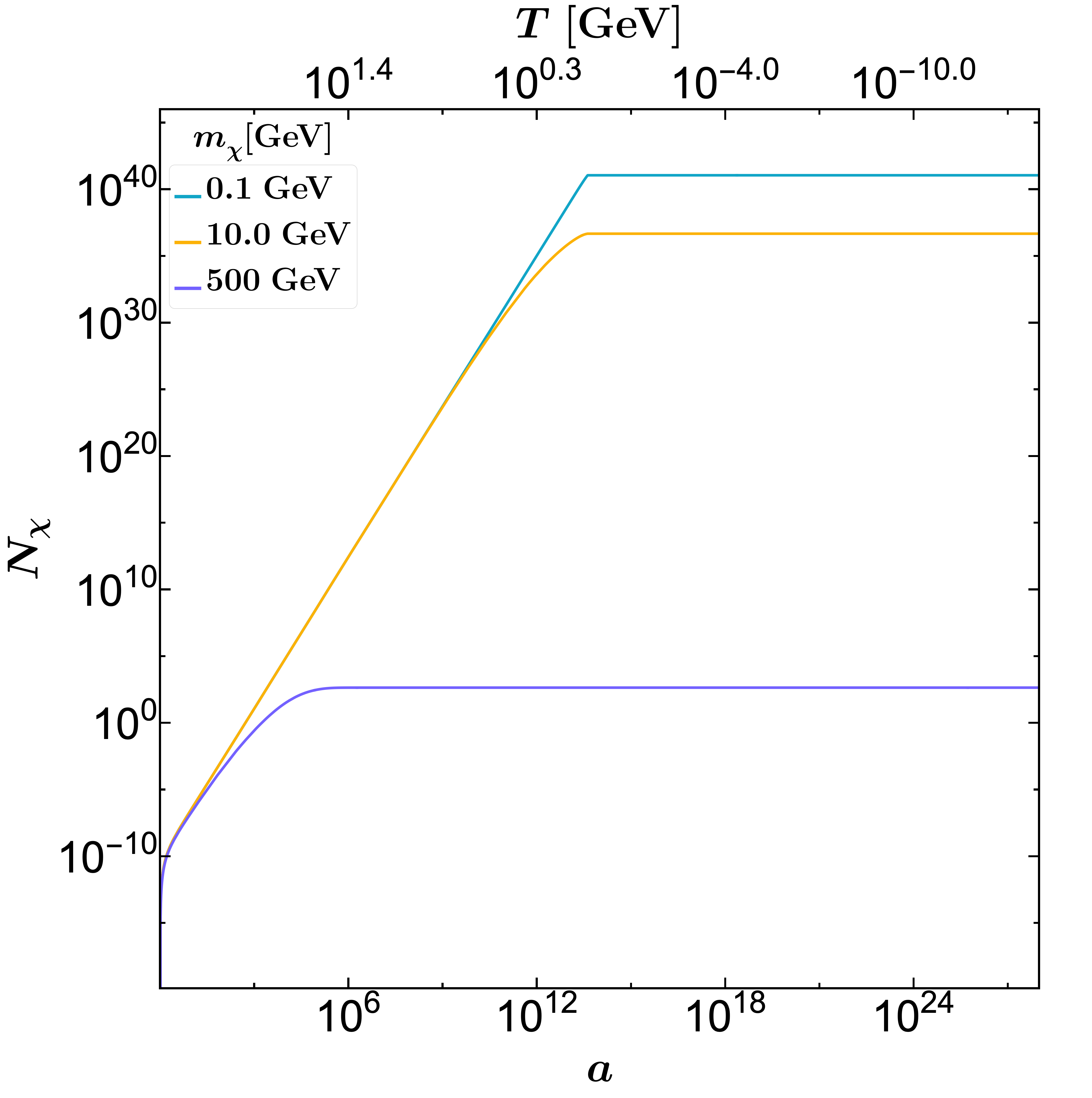}
\caption{Figures represent the evaluation of co-moving fermionic DM number density ($N_{\rm DM}=\ndm a^3$) with the scale factor ($a$). Here, we assumed all other Wilson coefficients are unity and fixed other parameters in the plots as: $n=6,$ $[\lNP,~\trh]=[1.0 ~\tev,0.1~\gev]$ (left) and $[10 ~\tev,1~\gev]$ (right).}
\label{fig:gammaH-N2}
\end{figure}

\noindent Fig. \ref{fig:gammaH-N} shows the evolution of the comoving number density of DM with scale factor for fixed $\lNP, m_{\rm DM}$, and $n=6$. Dashed and solid contours correspond to equilibrium and freeze-in abundances, respectively. Fig. \ref{fig:gammaH-N2} shows fermionic DM yield for different masses and $n=6$. For very large DM mass $m_{\rm DM} \gg \Tmax$, the yield remains negligible due to Boltzmann suppression throughout the DM production era.
\section{Signal-to-noise ratio for GW}
\label{sec:SNR}

For a given amplitude of the stochastic gravitational wave signal ($\Omega_{\rm GW}(f)$), the signal-to-noise ratio (SNR) of the signal for a given experiment is calculated as \cite{Maggiore:1999vm, Allen:1997ad} 
\begin{eqnarray}
    \text{SNR} = \left[\tau_{\rm obs} \int_{f_{\rm min}}^{f_{\rm max}} \left(\frac{\Omega_{\rm GW}(f)h^2}{\Omega_{\rm noise}(f)h^2}\right)^{2}df\right]^{1/2},
\end{eqnarray}
where $\Omega_{\rm noise}(f)$ represents the detector noise energy density and $f_{\rm min}$ ($f_{\rm max}$) denotes the minimum (maximum) operational frequency range of the given GW detector. The detector noise energy density $\Omega_{\rm noise}(f)$ is related to the noise characteristic strain $h_{c}(f)$ by the relation $\Omega_{\rm noise}(f)=\frac{2\pi^2 f^2}{3 H_{0} }h^2_{c}(f)$. The term $\tau_{\rm obs}$ represents the observational time of the particular GW detector. In the calculations, we use $10$ years of observational time, $\tau_{\rm obs}$ for all GW detectors.

\bibliographystyle{JHEP} 
\bibliography{ref.bib, ref-eft.bib}
\end{document}